\documentclass[pra,onecolumn,floatfix,a4paper,superscriptaddress]{revtex4}
\usepackage{bm,color,graphicx,amsmath,txfonts}

\usepackage[colorlinks, citecolor=blue,linkcolor=blue]{hyperref}
\begin{document}

	\title{Generating three transparency windows, Fano-resononce and slow/fast light in magnomechanical system through an auxiliary microwave cavity}

	\author{M'bark Amghar} 
	\affiliation{LPTHE-Department of Physics, Faculty of sciences, Ibnou Zohr University, Agadir, Morocco}
	
	\author{Chabar Noura}
	\affiliation{LPTHE-Department of Physics, Faculty of sciences, Ibnou Zohr University, Agadir, Morocco}
	
	\author{Mohamed Amazioug}\email{m.amazioug@uiz.ac.ma}
	\affiliation{LPTHE-Department of Physics, Faculty of sciences, Ibnou Zohr University, Agadir, Morocco}
	
	\author{Amjad Sohail Shah}
	\affiliation{Department of Physics, Government College University, Allama Iqbal Road, Faisalabad 38000, Pakistan}
	\affiliation{Instituto de F\'isica Gleb Wataghin, Universidade Estadual de Campinas, Campinas, SP, Brazil}
	
	\begin{abstract}
		
		In this paper, we propose a theoretical scheme to investigate the magnomechanically induced transparency (MMIT) phenomenon, Fano resonances, and slow/fast light effects in a hybrid cavity magnomechanical system. The magnomechanical system consists of two cavities: the principal cavity contains two ferromagnetic yttrium iron garnet (YIG) spheres, and the auxiliary cavity contains an atomic assembly. These two cavities are connected via photon tunneling, with the principal cavity being driven by two electromagnetic fields. The photon-magnon and phonon-magnon couplings are responsible for the magnon-induced transparency (MIT) and MMIT observed in the probe output spectrum. Furthermore, we examine the impacts of tunneling coupling, atom-photon coupling, and the magnetic field on the absorption, dispersion, and transmission spectra. We provide an explanation of the mechanism behind the Fano resonance phenomenon. Additionally, we address the phenomenon of slow and light propagation. Moreover, we demonstrate that group delay of the probe field can be improved by increasing photon tunneling strength. We also show that the slow light profile is decreased by adjusting the atom-photon coupling strength. This model is experimentally feasible. We hope these findings have the potential to be applied to the processing of quantum information and communication.
		
	\end{abstract}
	
	\date{\today}
	
	\maketitle
	\section{INTRODUCTION}
	Electromagnetically induced transparency (EIT) \cite{1,2,3,4} is a quantum interference phenomenon observed in multi-level atomic systems. It arises when these systems are subjected to intense laser fields, resulting in the suppression of absorption for a weak probe light. Initially predicted \cite{2} by Harris et al. and experimentally demonstrated \cite{3} by Boller et al. using strontium vapor, EIT has become a widely researched phenomenon. This research has unveiled various novel occurrences, such as quantum memories for photons \cite{5,6}, slow light \cite{7,8}, enhanced nonlinearity \cite{9}, vibrational cooling \cite{10}, optical switch \cite{11} and multiple EITs with two or more transparency windows \cite{12,13,14,15,16}. Typically, multiple EITs arise in systems with coherent multi-channels, enabling prolonged nonlinear interactions between weak fields, and have been proposed as a method to coherently control photon-photon interactions \cite{17}.\\
	Optomechanically induced transparency (OMIT), a phenomenon analogous to EIT, was first explored in 2010 by Weis et al. within vibrational cavity optomechanical systems \cite{18}. This effect emerges due to destructive interference among various internal field pathways in the system, creating a transparency window for probe light in spectral regions that would typically exhibit strong absorption \cite{19}. This phenomenon is proposed for use in various applications, ultra slow light propagation \cite{20}, including quantum routers \cite{21}, precision measurement \cite{22} and four-wave mixing \cite{23}. So far, the emergence of various new cavity optomechanical systems has led to theoretical research on the multiple-OMIT phenomenon. This research includes multiple-resonator optomechanical systems \cite{26}, hybrid piezo-optomechanical cavity systems \cite{24,244} and atomic-media assisted optomechanical systems \cite{25,Tesfay2021}, single-photon emitter \cite{Abdi2019}, Bose–Einstein condensate \cite{Asjad2013}, nonlinear optomechanical system \cite{Shahidani2013}, among others.\\

Recently, the cavity magnonic system (CMS) featuring a yttrium iron garnet (YIG) sphere has garnered significant interest, much like the field of cavity optomechanics. This system offers numerous advantages over traditional systems due to the unique properties of magnetic materials, such as low damping rates, high spin density, and strong cooperativity with microwave photons \cite{27,28,29,30,31,32}. These characteristics make CMS an ideal platform for studying strong light-matter interactions. Moreover, the magnon has excellent coupling capabilities with various quantum systems, including optical photons, microwaves\cite{33}, optical whispering gallery modes (WGMs) \cite{34}, superconducting qubits \cite{35}, and  phonons \cite{36}. Due to these characteristics, YIG systems are employed to investigate a variety of coherent phenomena similar to those observed in optomechanical systems. These phenomena include ground-state cooling of magnomechanical resonators \cite{37}, quantum entanglement \cite{38,381,39,40,}, magnomechanically induced transparency (MMIT) \cite{41,411,42,43}, as well as the engineering of slow and fast light \cite{42,43,44,444,4444}, and the phenomena of bistability and nonreciprocity \cite{45,46}.\\
	In their CMS research, Ullah $et$ $al$, explored various intriguing quantum effects by exploiting the nonlinearity of magnetostrictive interactions in a system of two ferromagnetic YIG spheres coupled to a single microwave cavity mode \cite{43}. They have demonstrated the MMIT and magnon induced transparency (MIT) \cite{42,43,47}. In addition, they have demonstrated the emergence of the Fano resonance phenomenon of the probe field \cite{48,49,50}. They have also demonstrated the phenomenon of slow/light propagation. Liao $et$ $al$ have studied MMIT as well as slow and fast light effects in a system comprising a high-quality yttrium-iron garnet (YIG) sphere and an atomic ensemble, both located inside a microwave cavity \cite{42}. Recently, Liu $et$ $al$. have suggested a scheme for an auxiliary-microwave-assisted coupled magnomechanical model (AMCA) involving two yttrium-iron garnet (YIG) spheres. They have introduced a mechanism for simultaneously enhancing quantum cooling and entanglement by coupling an auxiliary microwave cavity to a magnomechanical cavity \cite{51}.\\
	In this paper, we investigate the MMIT phenomenon, Fano resonance, transmission and slow/fast light effect in a magnomechanical system in which consists of two single-cavities connected by single photon hopping, encompassing an atomic ensemble and two yttrium-iron-garnet (YIG) spheres, see Fig. \ref{1}. We investigate the effect of the photon hopping coupling strength, atom-photon coupling strength and the magnetic field on the absorption and dispersion spectra. Moreover, we study the Fano resonances in the output probe field. Additionally, we explore the phenomenon of slow and fast light propagation, clarifying that the group delay is influenced by the tunability of the photon hopping coupling strength, as well as the atom-photon coupling.\\
	The organization of this paper is delineated as follows: Sect. \ref{0} introduces the system, presents the formula for its Hamiltonian, and outlines the relevant quantum Langevin equations (QLEs) while calculating the output field. In Sect. \ref{01}, we examine magnomechanically induced transparency and analyze how the photon hopping coupling strength $f$, atom-photon coupling strength and magnetic field affect the input spectrum. Sect. \ref{FFF} focuses on the study of Fano resonance in the output field. Sect. \ref{004} describes the transmission of the probe field and evaluates the group delays related to slow and fast light propagation. In Sect. \ref{00} we examine the feasibility of our system using current experimental parameters and configurations. Finally, the paper concludes with concluding remarks.

	\section{MODEL} \label{0}
\begin{figure}[t]
		\centering
		\hskip-1.0cm\includegraphics[width=0.9\linewidth]{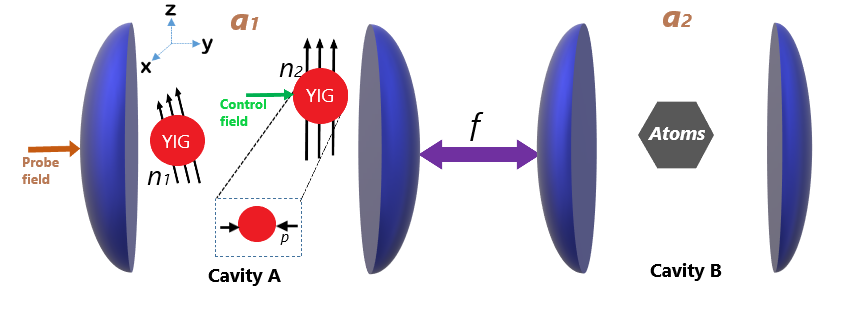}
		\caption{A schematic representation of two single-mode cavities, designated as A and B, which are interconnected with a coupling strength denoted by $f$. Cavity A encompasses two YIG spheres and is stimulated by an external laser field (probe field) operating at a frequency denoted as $\omega_d$. Each sphere is positioned within a uniform bias magnetic field and near the peak magnetic field of the magnomechanical cavity, facilitating the interaction between the photon mode $a_1$ and the two magnon modes, $n_1$ and $n_2$. The right sphere is directly affected by an external microwave field oriented in the \textsf{y}-direction, which functions as a control field within our model. Cavity B houses an ensemble of N two-level atoms, each defined by an intrinsic frequency denoted as $\omega_u$.}\label{1}
	\end{figure}

	In the hybrid coupled-cavity magnomechanical system depicted in Fig. \ref{1}, we consider two single-mode cavities connected by single-photon hopping $\mathit{f}$, each with resonance frequencies $\omega_r$ $(r = 1, 2)$, including an atomic ensemble and two YIG spheres. This interconnected system includes six excitation modes: microwave electromagnetic modes in cavity A and cavity B, phonon and two magnon modes in the YIG sphere, and atomic excitation within cavity B.\\
	In cavity A, each sphere is subjected to a uniform magnetic field in the \textsf{z} direction, which stimulates magnonic modes. These modes are interconnected with the cavity A field by dipolar magnetic interactions. Excitation of the magnonic modes inside the sphere results in fluctuating magnetization, which induces changes in the lattice structure. The magnetostrictive force generates vibrations in the yttrium iron garnet (YIG) spheres, leading to interactions between the magnons and phonons inside these spheres. The strength of the single-magnon magnomechanical coupling is notably feeble \cite{49} and is contingent upon both the diameter of the sphere and the orientation of the external polarization field. By considering a larger $n_1$ sphere, or by changing the orientation of the polarizing magnetic field applied to it, the magnetomechanical coupling of this sphere can be neglected \cite{40}. In this context, we suppose that the orientation of the polarizing field on the $n_1$ sphere is configured to make the magnomechanical single-magnon interaction extremely weak, allowing it to be ignored \cite{49}. Nevertheless, the magnomechanical interaction of $n_2$ sphere is significantly augmented by the direct excitation of its magnon mode through the application of an external microwave actuator. This microwave driver functions as the control field within our model. Furthermore, cavity A is stimulated by a feeble probe field.\\
	Within cavity B, an ensemble comprising $N$ two-level atoms, characterized by a transition frequency of $\omega_u$ interacts with the cavity field. Each atom in the ensemble is identified by the spin $-1/2$ Pauli matrices $\sigma_+$, $\sigma_-$, and $\sigma_z$. The collective spin operators associated with atomic polarization within the ensemble are represented as $S_{+,-,z}$ = $\sum_{i=1}^{N}$$\sigma^i_{+,-,z}$. These operators conform to the following commutation relations: $[S_+, S_-]=S_z$ and $[S_z, S_{\pm}] = {\pm}2S_{\pm}$ \cite{m3,m4}. The above operators $S_{\pm}$ and $S_z$ can be formulated in terms of bosonic annihilation and creation operators $u$ and $u^{\dagger}$, respectively, by employing the Holstein-Primakoff transformation \cite{m4,m44,m5,m6}: $S_+ = u^{\dagger}\sqrt{N - u^{\dagger}u} \approx u^{\dagger}\sqrt{N}$, $S_- = u\sqrt{N - u^{\dagger}u} \approx u\sqrt{N}$, $S_z = u^{\dagger}u - \frac{N}{2}$, where $u$ and $u^{\dagger}$ satisfy the commutation relation $[u, u^{\dagger}] = 1$. It is important to note that this transformation is only valid on condition that the population of atoms in the ground state considerably exceeds that of atoms in the excited state, ensuring that $S_z \approx \langle S_z \rangle \approx -N$ \cite{m7}. The complete Hamiltonian that characterizes the system can be expressed as follows
	\begin{equation}\label{e1}
		\begin{aligned}
			H / \hbar= & \sum_{r=1}^2 \omega_r a_r^{\dagger} a_r+\omega_u u^{\dagger} u+\sum_{r=1}^2 {\omega_n}_r n^{\dagger}_r n_r+{\omega_p}p^\dagger p +g_{n p} n^{\dagger}_2 n_2 \left(p+p^\dagger\right)+f\left(a_1^{\dagger} a_2+a_1 a_2^{\dagger}\right)+G_{a u}\left(u a_2^{\dagger}+u^{\dagger} a_2\right) \\
			& +\sum_{r=1}^2g_{r}\left(a_1 n^{\dagger}_r+a_1^{\dagger} n_r\right)+i\left( a_1^{\dagger}\epsilon_d e^{-i\omega_dt}-a_1\epsilon_d^*e^{i\omega_d t}\right)+i \Omega\left(n^{\dagger}_2 e^{-i \omega_{0 } t}-n_2 e^{i \omega_{0 } t}\right).
		\end{aligned}
	\end{equation}
	The first four terms of Eq. \eqref{e1} represent the energy associated with the cavities (A and B), atomic excitation, magnon (\( n_1 \) and \( n_2 \)), and phonon modes, respectively. Here, \( \omega_r \) describe the frequencies inside each cavity, \( \omega_u \) is the intrinsic frequency of the two-level atoms in the atomic ensemble, \( \omega_{nr} \) denotes the frequencies of each magnon, and \( \omega_p \) is the frequency of the phonon mode. Notably, \( \omega_{nr} \) can be modified through the manipulation of the external bias magnetic field, \( H_r \) as \( \omega_{nr} = \gamma H_r \). The operator $a_r(a_r^\dagger)$, $u(u^\dagger)$, $n_r(n_r^\dagger)$ and $p(p^\dagger)$ are the annihilation (creation) operators of the cavity, collective atomic excitation, magnon and phonon modes, respectively, adhere to the canonical commutation relations: \( [a_r, a_r^\dagger] = 1 \), \( [u, u^\dagger] = 1 \), \( [n_r, n_r^\dagger] = 1 \), and \( [p, p^\dagger] = 1 \), where \( r = 1,2 \). The following four terms describe the interactions between interconnected subsystems within the cavity system: the coupling between the magnon $n_2$ and phonon is characterized by $g_{np}$; the coupling between cavity A and cavity B is represented by $f$; the interaction between the collective atomic excitation and cavity B is described by $G_{au}$; and finally, the coupling between cavity A and the magnon modes is denoted by $g_r$. The coupling rate of collective atomic excitation with the cavity mode is given by $G_{au}=h\sqrt{N}$, where $h$ denotes the atom-cavity coupling strength defined as $h=\nu \sqrt{\omega_2 / 2 \hbar \epsilon_0 V}$, with $\nu$ being the dipole moment of atomic transition, $V$ the volume of the cavity, and $\epsilon_0$ the permittivity of free space. The second-last term in the Equation \eqref{e1} describes a microwave field driving cavity A with $\omega_d$ and $\epsilon_d=\sqrt{2 \kappa_{a} P_d / \hbar \omega_d}$ are, respectively, the frequency and the amplitude of the probe field, which depends on the input power $P_d$ of the probe field and the decay rate $\kappa_{a}$ of the cavity. Additionally, the last term represents a magnon-mode drive field with a Rabi frequency $\Omega=\frac{\sqrt{5}}{4} \gamma\sqrt{\text{N}} B$ , where $B$ represents the amplitude of the applied field, $\gamma=28$ $\text{GHz}/\text{T}$ indicates the gyromagnetic ratio and $\text{N}=\rho \mathcal{V}$ signifies the total number of spins with the spin density of the YIG sphere $\rho=4.22\times10^{27}\text{m}^{-3}$ and volume $\mathcal{V}$ \cite{m8}.\\
	In Eq. \eqref{e1}, it is important to note that we have disregarded the nonlinear term $Kn^\dagger_rn^\dagger_rn_rn_r$, which could potentially emerge from a strongly driven magnon mode \cite{46}. To neglect this nonlinear term, it is required that $K|\langle n_2\rangle|^3<<\Omega$. For the specific parameters of the system studied in this work, this condition is systematically respected. The Hamiltonian that characterizes the system within the framework of the rotating-wave approximation, in a reference frame that rotates in accordance with the drive field frequency \(\omega_0\), is articulated as follows
	\begin{equation} \label{e2}
		\begin{aligned}
			H / \hbar= & \sum_{r=1}^2 \Delta_r a_r^{\dagger} a_r+\Delta_u u^{\dagger} u+\sum_{r=1}^2 \Delta_{n_r} n^{\dagger}_r n_r+{\omega_p}p^\dagger p +g_{n p} n^{\dagger}_2 n_2 \left(p+p^\dagger\right)+f\left(a_1^{\dagger} a_2+a_1 a_2^{\dagger}\right)+G_{a u}\left(u a_2^{\dagger}+u^{\dagger} a_2\right) \\
			& +\sum_{r=1}^2g_{r}\left(a_1 n^{\dagger}_r+a_1^{\dagger} n_r\right)+i\left( a_1^{\dagger}\epsilon_d e^{-i\delta t}-a_1\epsilon_d^*e^{i\delta t}\right)+i \Omega\left(n^{\dagger}_2  -n_2  \right),
		\end{aligned}
	\end{equation}
	where $\Delta_r=\omega_r-\omega_{0 }$, $\Delta_u=\omega_u-\omega_{0 }$, $\Delta_{n_r}=\omega_{n_r}-\omega_{0 }$ and $\delta=\omega_d-\omega_{0 }$. The Heisenberg-Langevin quantum equations resulting from Hamiltonian \eqref{e2} can be stated as follows
	\begin{equation}
		\begin{aligned}
			& \dot{a}_1=-\left(\kappa_a+i \Delta_1\right) a_1-i g_{1}n_1-ig_2n_2-i f a_2+\epsilon_de^{-i\delta t}+\sqrt{2 \kappa_a} a_1^{i n}, \\
			& \dot{a}_2=-\left(\kappa_a+i \Delta_2\right) a_2-i f a_1-i G_{au} u+\sqrt{2 \kappa_a} a_2^{i n}, \\
			& \dot{p}=-\left(\kappa_p+i \omega_p\right) p-i g_{np} n^\dagger_2 n_2 +\sqrt{2 \kappa_p} p^{i n}, \\
			& \dot{u}=-\left(\gamma_u+i \Delta_u\right) u-i G_{a u} a_2+\sqrt{2 \gamma_u} u^{i n}, \\
			& \dot{n_1}=-\left(\kappa_{n_1}+i \Delta_{n_1}\right) n_1-i g_{1} a_1+\sqrt{2 \kappa_{n_1}} n_1^{i n}, \\
			& \dot{n_2}=-\left(\kappa_{n_2}+i {\Delta_{n_2}}\right) n_2-i g_{2} a_1-i g_{n p} n_2 \left(p^\dagger+p\right)+\Omega+\sqrt{2 \kappa_{n_2}} n_2^{i n}, \\
		\end{aligned}
	\end{equation}
    where $\kappa_{a}$, $\gamma_u$, and $\kappa_{n_r}$ denote the decay rates associated with the cavities (A and B), the atomic system, and the magnon modes, respectively. $\kappa_p$ denotes the dissipation rate associated with the phonon mode. $a_r^{i n}$, $p^{i n}$, $u^{i n}$, and $r^{i n}$ represent the input noise operators in the vacuum, each with zero mean value, i.e., $\langle c^{in} \rangle = 0$ $(c = a_1,a_2, n_1,n_2, u, p)$ \cite{m10}. The magnon mode $n_2$ is strongly influenced by a microwave drive, resulting in a large steady-state amplitude $(|\langle n_{2s}\rangle| \gg 1)$ for the magnon mode. This interaction, along with the beam splitter, also leads to a substantial steady-state amplitude $(|\langle a_{1s}\rangle| \gg1)$ for the cavity  mode. As a result, we are capable of linearize the quantum Langevin equations in relation to the steady-state values, concentrating exclusively on the first-order terms associated with the fluctuating operator: $\langle \mathcal{R}\rangle=\mathcal{R}_{s}+\mathcal{R}_{-} e^{-i \delta t} +\mathcal{R}_{+}e^{i \delta t}$, where $\mathcal{R}= a_1,a_2, n_1,n_2, u, p$ \cite{m11}. We shall commence our analysis with the zero-order solution, which is associated with the steady-state solutions
	\begin{equation}\label{0551}
		\begin{aligned}
			&  a_{1s}=\frac{-ig_1n_{1s}-ig_2n_{2s}-i f a_{2s}}{\kappa_a+i\Delta_1}, \quad a_{2s}=\frac{-i f a_{1s}-iG_{au}u_s}{\kappa_a+i\Delta_2}, \\	
			& p_s=\frac{-ig_{n p}|n_{2s}|^2}{\kappa_p+i\omega_p}, \quad u_s=\frac{-i G_{a u} a_{2s}}{\gamma_u+i \Delta_u},\\						
			& n_{1s}=\frac{-i g_1 a_{1s}}{i \Delta_{n_1}+\kappa_{n_1}}, \quad n_{2s}=\frac{\Omega-i g_{2} a_{1s}}{i \tilde{\Delta}_{n_2}+\kappa_{n_2}}.\\		
		\end{aligned}
	\end{equation}

Assuming that the coupling between the external microwave drive and the magnon mode $ (n_2) $ is significantly stronger than the amplitude $ \epsilon_d $ of the probe field, we can solve the linearized quantum Langevin equations by focusing on the first-order perturbed solutions and neglecting all higher-order terms of $ \epsilon_d $. The cavity mode solution can then be expressed as follows
	\begin{equation}
		a_{1-}=\epsilon_d.\left( \mathcal{S}_1+\frac{f^2}{\mathcal{S}_{10}\mathcal{W}_6}+\frac{g_1^2}{\mathcal{S}_{12}}+\frac{g_2^2}{\mathcal{S}_{2}\mathcal{W}_5}\right)^{-1}, 
	\end{equation} 	
	where 
	$$\mathcal{S}_1=\kappa_a+i(\Delta_1-\delta),\quad \mathcal{S}_2=\kappa_{n_2}+i(\tilde{\Delta}_{n_2}-\delta) ,\quad \mathcal{S}_3=\kappa_{p}+i(\omega_{p}-\delta),\quad \mathcal{S}_4=\kappa_{p}-i(\omega_{p}+\delta),\quad \mathcal{X}=\left(1-\mathcal{S}_3/\mathcal{S}_4\right),$$	
	$$\mathcal{S}_5=\kappa_{n_2}-i(\tilde{\Delta}_{n_2}+\delta) ,\quad \mathcal{S}_6=\kappa_{a}-i(\Delta_{1}+\delta),\quad
	\mathcal{S}_7=\kappa_{a}-i(\Delta_{2}+\delta),\quad
	\mathcal{S}_8=\gamma_{u}-i(\Delta_{u}+\delta),\quad
	\mathcal{S}_9=\kappa_{n_1}-i(\Delta_{n_1}+\delta),$$
	$$\mathcal{S}_{10}=\kappa_{a}+i(\Delta_{2}-\delta),\quad
	\mathcal{S}_{11}=\gamma_{u}+i(\Delta_{u}-\delta),\quad
	\mathcal{S}_{12}=\kappa_{n_1}+i(\Delta_{n_1}-\delta),\quad
	\mathcal{W}_1=1+\frac{G_{au}^2}{\mathcal{S}_7\mathcal{S}_8},\quad \mathcal{W}_2=1+\frac{g_{1}^2}{\mathcal{S}_6\mathcal{S}_9}+\frac{f^2}{\mathcal{S}_6\mathcal{S}_7\mathcal{W}_1},$$                                                 	
	$$\mathcal{W}_3=1+\frac{g_{2}^2}{\mathcal{S}_5\mathcal{S}_6\mathcal{W}_2},\quad
	\mathcal{W}_4=1-\frac{G_{A}^2\mathcal{X}}{\mathcal{S}_3\mathcal{S}_5\mathcal{W}_3},\quad
	\mathcal{W}_5=1+\frac{G_{A}^2\mathcal{X}}{\mathcal{S}_2\mathcal{S}_3\mathcal{W}_4},\quad
	\mathcal{W}_6=1+\frac{G_{au}^2}{\mathcal{S}_{10}\mathcal{S}_{11}}.$$  
	Here $G_A=\frac{G_{np}}{\sqrt{2}}$, with $G_{np}=i\sqrt{2}g_{np}n_{2s}$ is the effective magnomechanical coupling. Utilizing the input-output standard relation for the cavity A: $\epsilon_{out}=\epsilon_{int}-2\kappa_{a}\langle a_1\rangle$ \cite{m12}, we can express the amplitude of the output field as	
	\begin{equation}
		\epsilon_{out}=\frac{2\kappa_a a_{1-}}{\epsilon_d}.
	\end{equation}
    The real part of the output probe field Re$[\epsilon_{out}]$ represents the absorption spectrum. The imaginary part of the output probe Im$[\epsilon_{out}]$ represents the dispersion spectrum.
    
    The steady-state Eq. \eqref{0551} can be expressed as follows
    \begin{equation}
    	n_{2s}=\frac{\mathcal{B}\Omega}{\mathcal{B}(\kappa_{n_ 2}+i\tilde{\Delta}_{n_2})+\mathcal{A}g_2^2(\kappa_{n_ 1}+i\Delta_{n_1})}
    \end{equation}
    Where
    $$\mathcal{A}=(\kappa_{a}+i\Delta_{2})(\gamma_{u}+i\Delta_{u})+G_{au}^2$$
     $$\mathcal{B}=\mathcal{A}(\kappa_{a}+i\Delta_{1})(\kappa_{n_ 1}+i\Delta_{n_1})+g_1^2\mathcal{A}+f^2(\gamma_{u}+i\Delta_{u})(\kappa_{n_ 1}+i\Delta_{n_1})$$
    \begin{figure} [h!] 
    	\begin{center}
    		\includegraphics[scale=0.4]{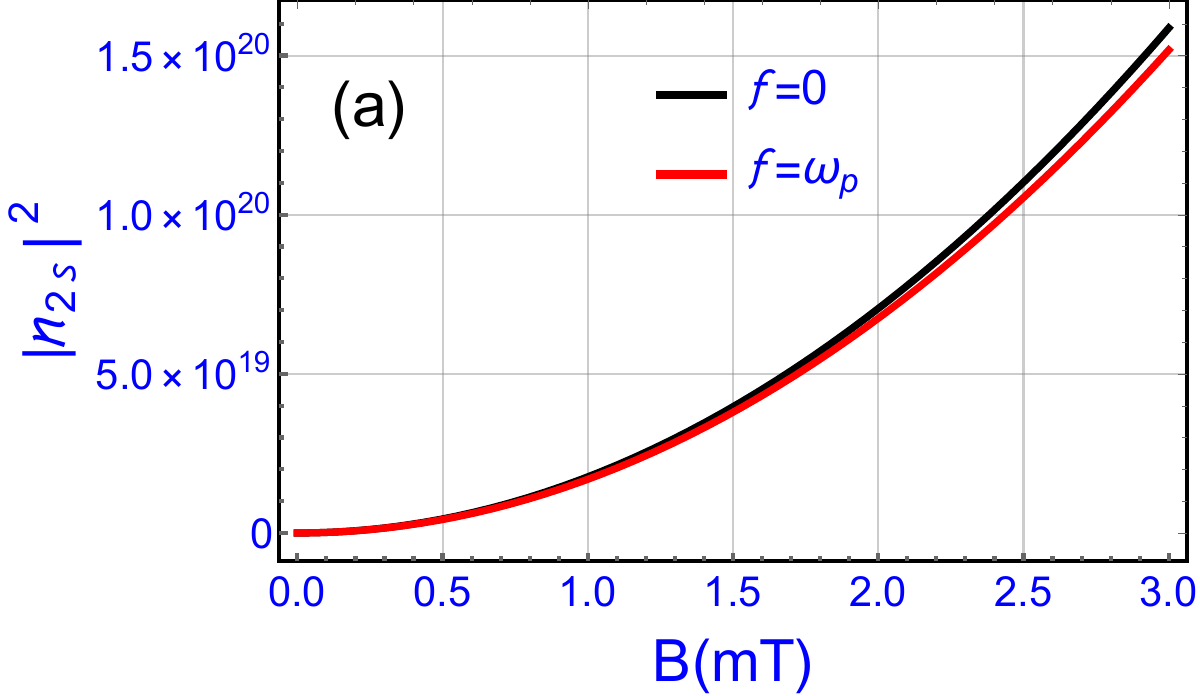}
    		\includegraphics[scale=0.4]{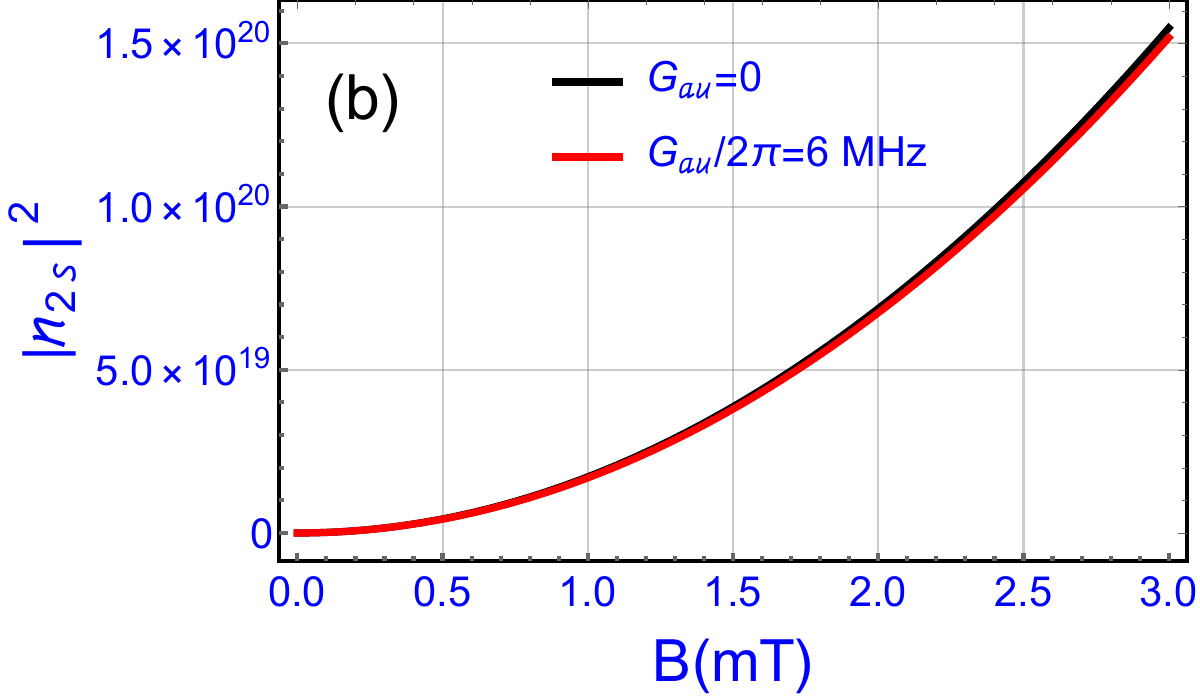}
    		\caption{Plot of steady-state magnon number as a function of applied magnetic field for different values of the tunneling coupling strength $f$ and the atom-photon coupling $G_{au}$ with $g_1/2\pi=1.2$ MHz and $g_1/2\pi=1.2$ MHz.}\label{NNN}
    	\end{center}
    \end{figure}
In Fig. \ref{NNN}(a), we plot the steady-state magnon number as a function of the driving magnetic field $B$ under the influence of the tunneling coupling strength. It can be seen that the number of spins increases with an increasing magnetic field $B$. We also observe that when the value of $B$ is fixed, the number of magnon in the stationary state decreases with increasing coupling intensity $f$. In Fig. \ref{NNN}(b), we plot the steady-state magnon number as a function of the driving magnetic field $B$ under the influence of the atom-photon coupling strength. We remark that the magnon number decreases with increasing the atom-photon coupling $G_{au}$ when the value of $B$ fixed.   
	\section{MAGNOMECHANICALLY INDUCED TRANSPARENCY} \label{01}
	In this section, we numerically examine the influence of coupling strength $f$, atom-photon coupling strength $G_{au}$, and the magnetic field $B$ on magnomechanically induced transparency within a double cavity magnomechanical system. We utilize the effective parameters derived from a recent experiment in the cavity magnomechanical system, such as \cite{m8}: $\omega_{1,2} / 2 \pi=10$ $\mathrm{GHz}$, $\omega_p / 2 \pi=10$ $\mathrm{MHz}, $ $\kappa_p / 2 \pi=$ $100 \mathrm{~Hz}$, $\omega_{n_1,n_2} / 2 \pi=10$ $\mathrm{GHz},$ $\kappa_a / 2 \pi=2.1$ $\mathrm{MHz},$ $\kappa_{n_ 1} / 2 \pi=$ $\kappa_{n_ 2} / 2 \pi=0.1$ $\mathrm{MHz},$ $g_1 / 2 \pi=g_2 / 2 \pi=1.5$ $\mathrm{MHz},$ $G_{n p} / 2 \pi=$ 3.5 $\mathrm{MHz}, \Delta_{1,2}=\omega_p,$ $\Delta_{n_1,n_2}=\omega_p$, and $\omega_0 / 2 \pi=10$ $\mathrm{GHz}$. Accordingly, the atom-cavity coupling and atomic decay rate are on the order of megahertz, with $ G_{au}/2\pi = 6$ $\text{MHz} $ and $ \gamma_u/2\pi = 1$ $\text{MHz} $ \cite{m13}. Moreover, the hopping rate $ f $ between the cavities is also on the order of megahertz.\\
	\begin{figure} [h!] 
		\begin{center}
			\includegraphics[scale=0.25]{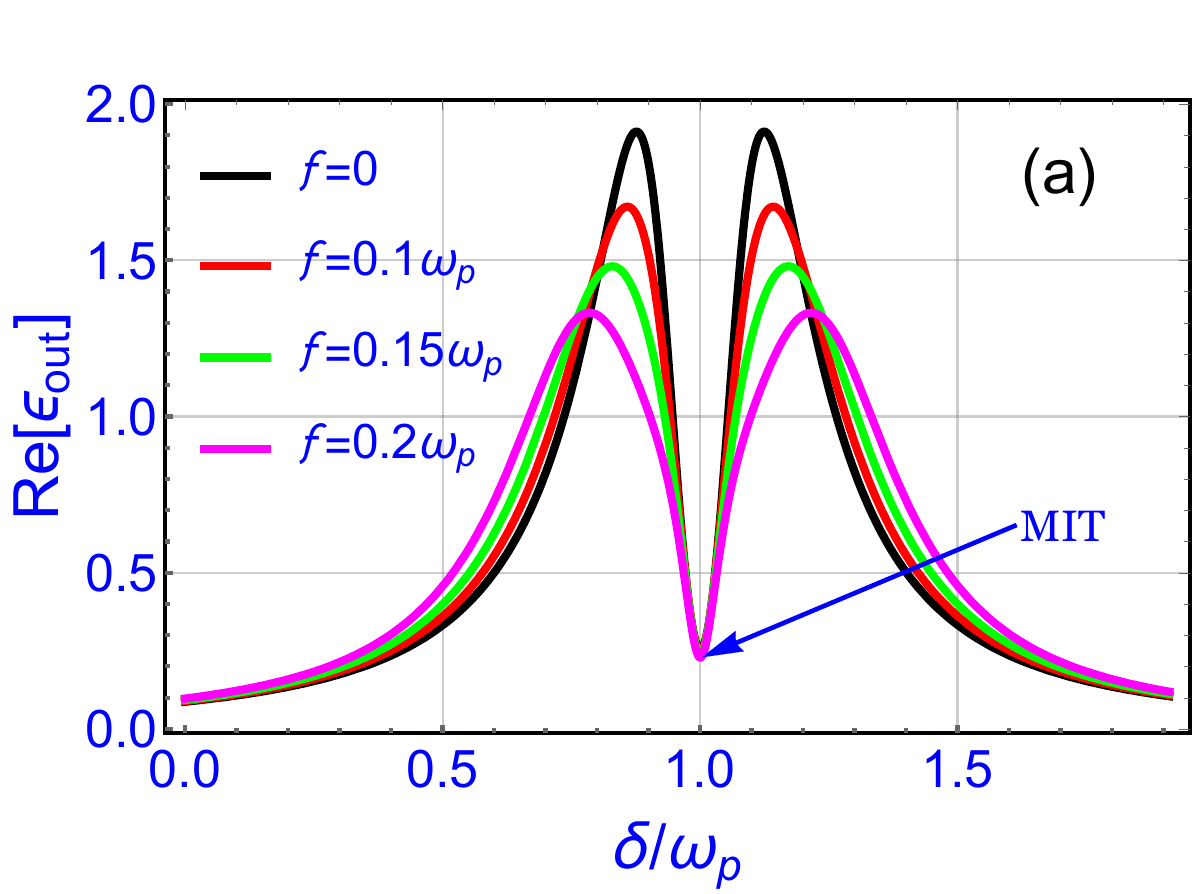}
			\includegraphics[scale=0.25]{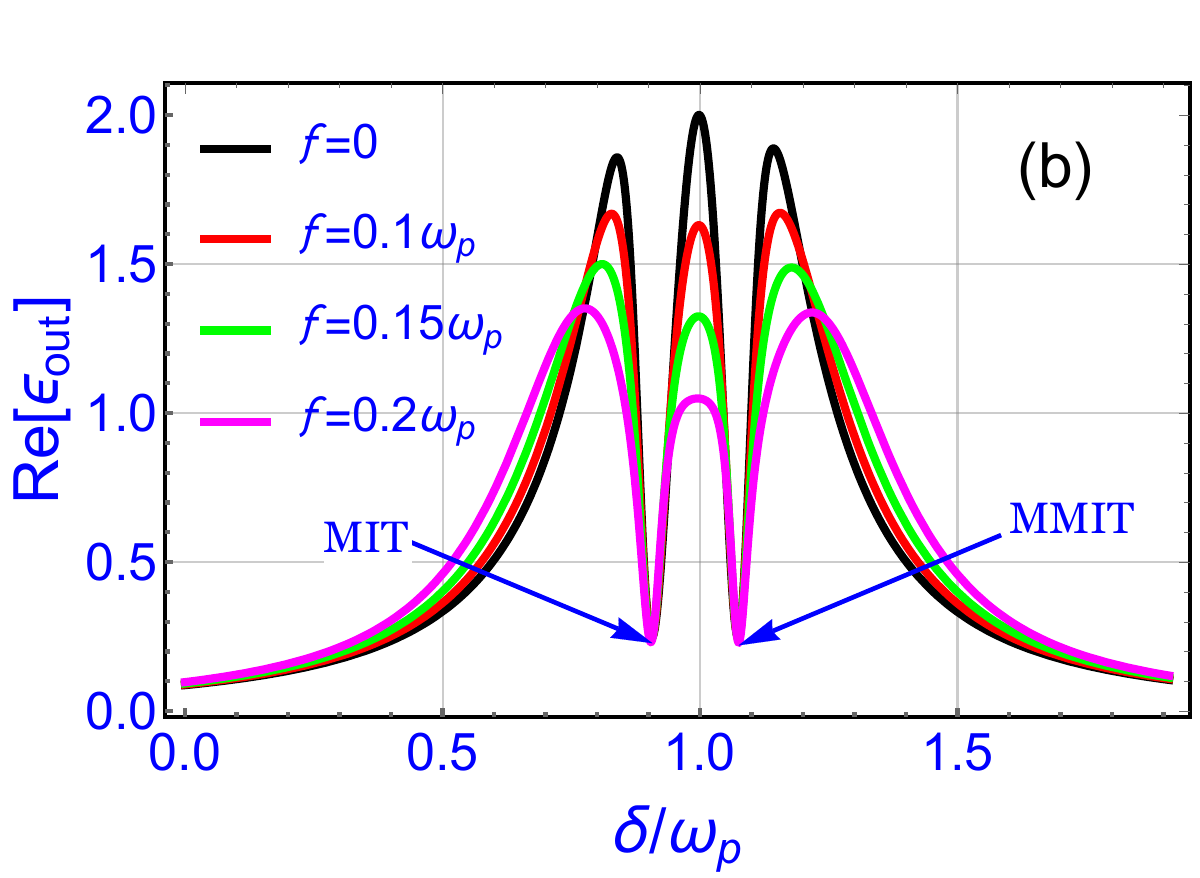}
			\includegraphics[scale=0.25]{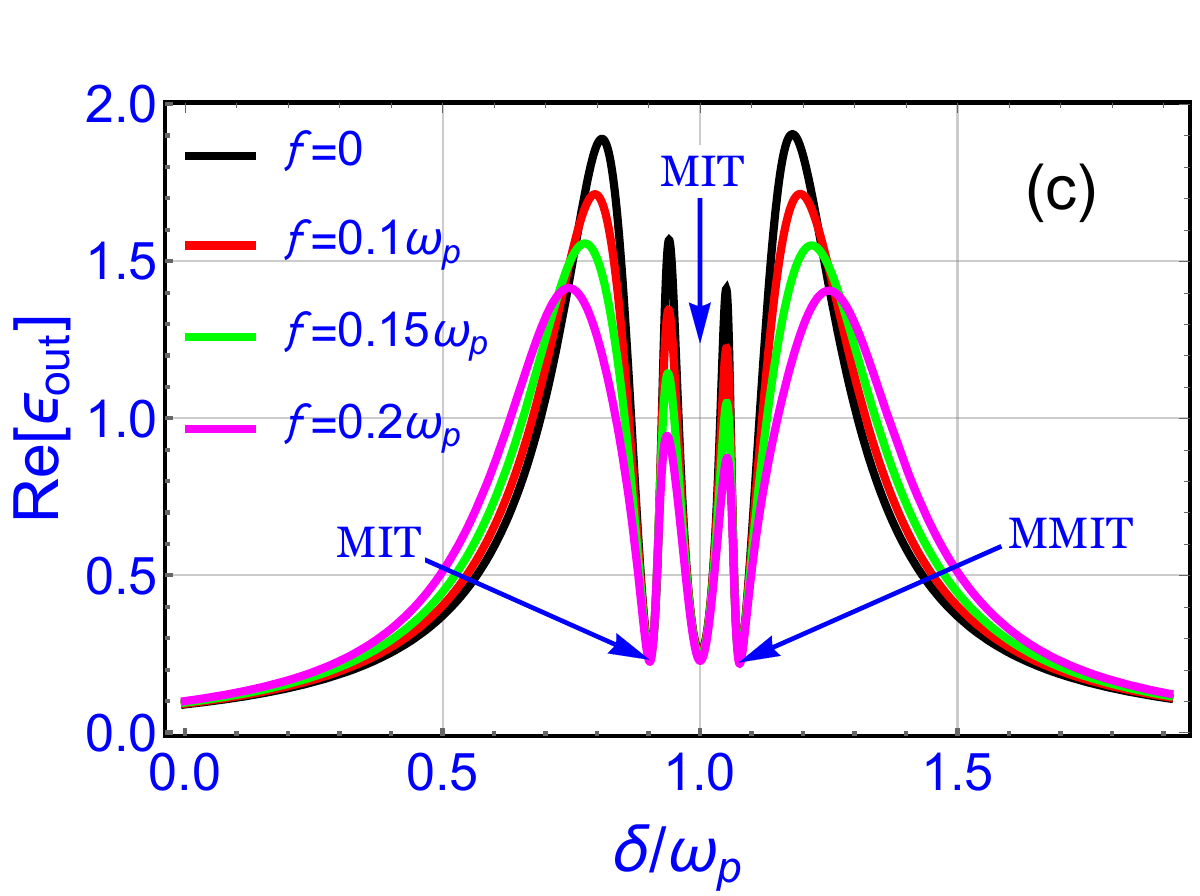}
			\includegraphics[scale=0.25]{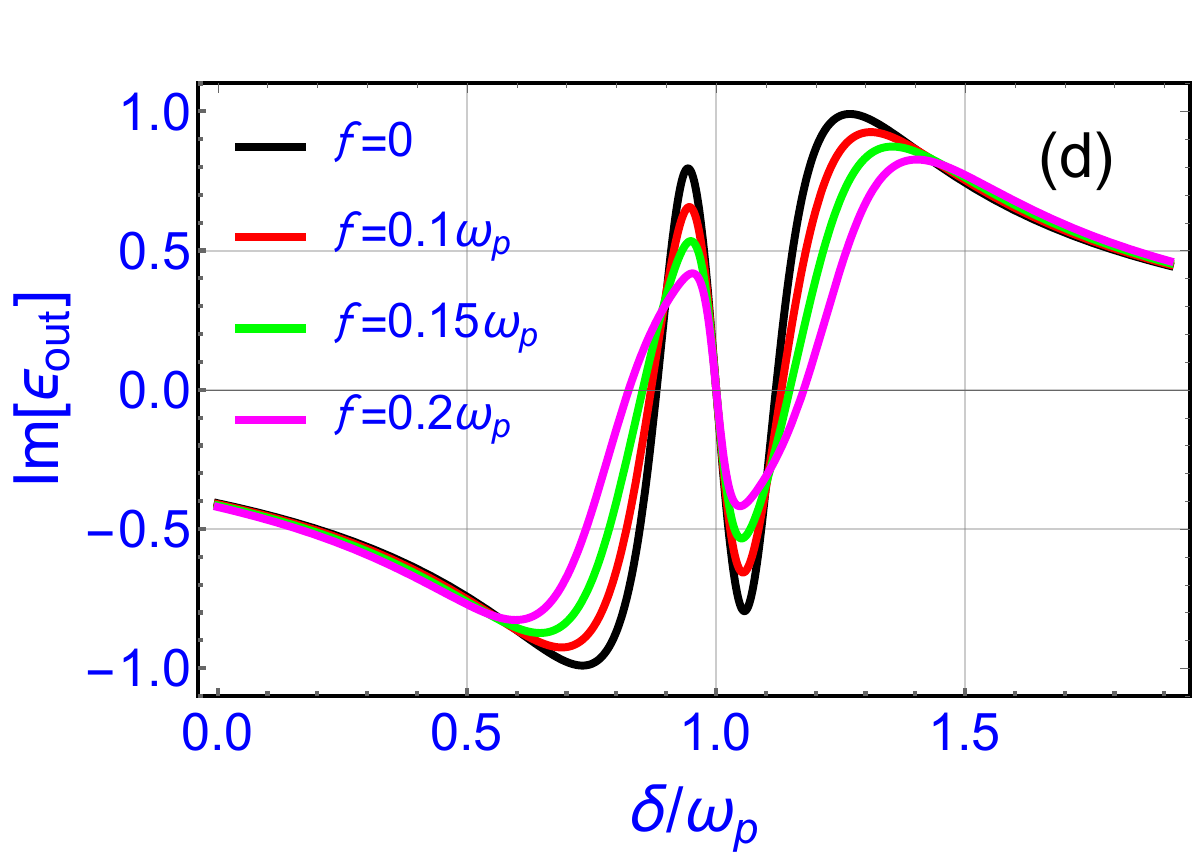}
			\includegraphics[scale=0.25]{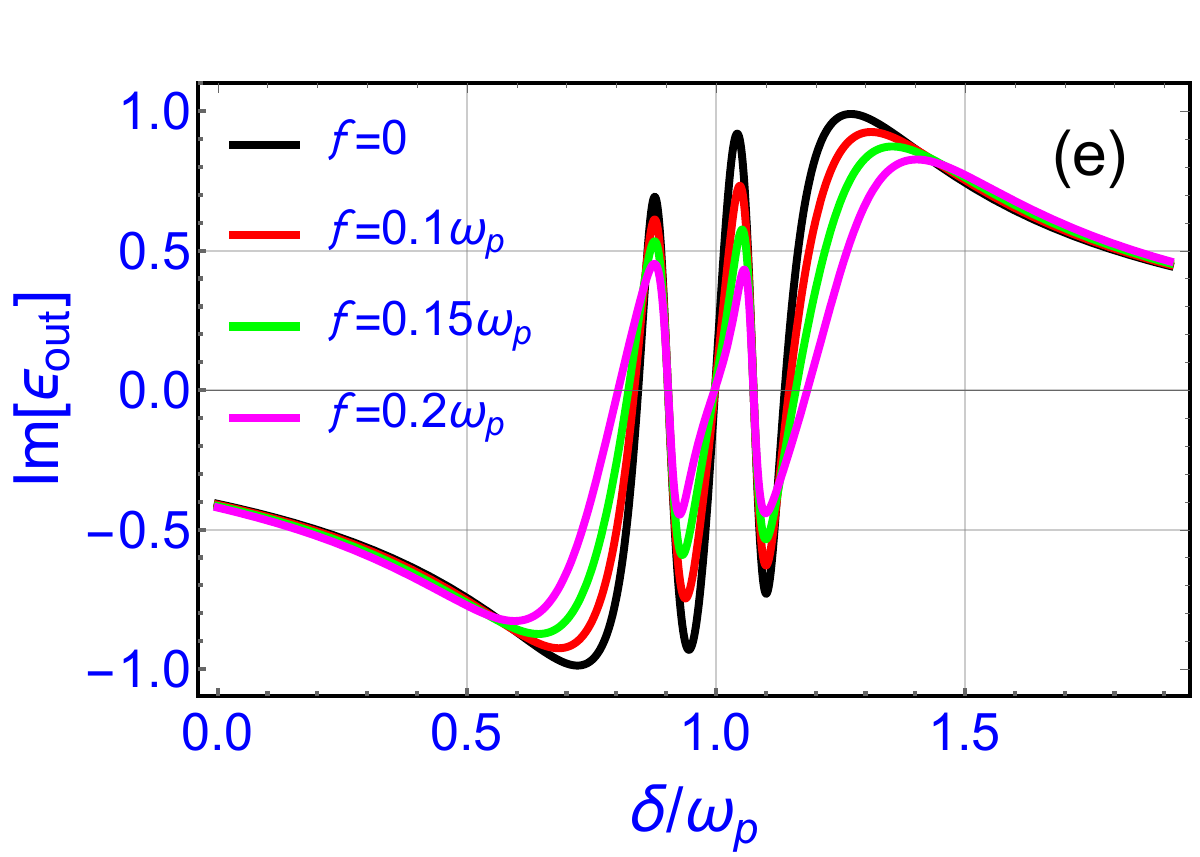}
			\includegraphics[scale=0.25]{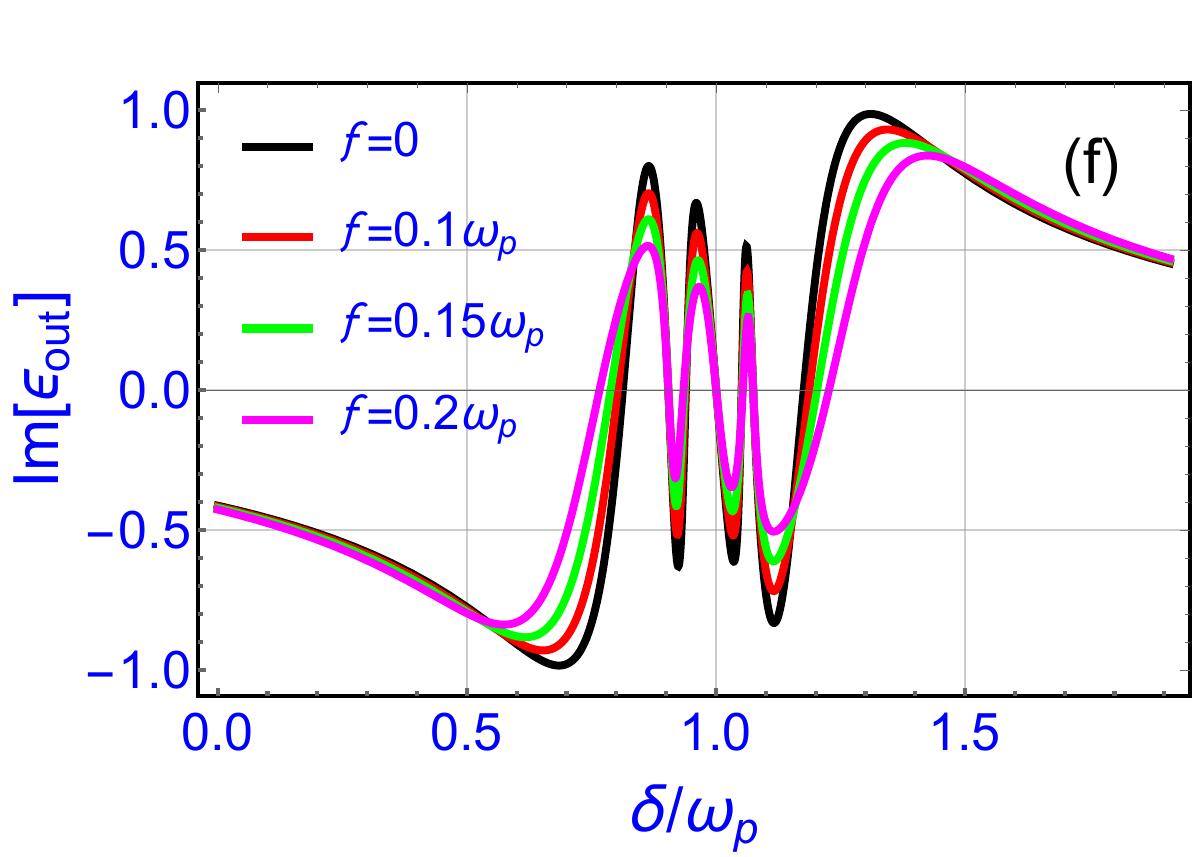}
			\caption{Plot of the real Re[$\epsilon_{out}$] and imaginary Im[$\epsilon_{out}$] parts of the output field as a function of the normalized detuning $\delta/\omega_{p}$ for different values of the tunneling coupling strength $f$ with $G_{au}=0$. (a)-(d) $g_1=G_{np}=0$ and $g_{2}/2\pi=1.2$ MHz, (b)-(e) $g_1=0$ and $G_{np}/2\pi=g_{2}/2\pi=1.2$ MHz, (c)-(f) $g_1/2\pi=g_{2}/2\pi=G_{np}/2\pi=1.2$ MHz. For additional parameters, see the text.} \label{b}
		\end{center}
	\end{figure} 
	In Figs. \ref{b}(a)-\ref{b}(c), we display the absorption spectrum Re[$\epsilon_{out}$] of the probe field as a function of the normalized probe detuning $\delta/\omega_{p}$ for different values of the cavity-cavity tunneling coupling $f$ and coupling strengths $g_1$, $g_2$ and $G_{np}$ with atom-cavity coupling is zero $(G_{au}=0)$. We suppose that the magnon-phonon coupling ($G_{np}$) and magnon-photon coupling ($g_1$) are disabled in Fig. \ref{b}(a).  As a result, cavity A is coupled only with magnon mode $ n_2 $. Based on these assumptions, we observe a single transparency window, called a magnon-induced transparency (MIT) window.  Double transparency windows are observed in the absorption spectrum when magnon-phonon coupling ($G_{np}$) is activated and $g_1$ are maintained at zero. The second transparency window associated with magnon-phonon coupling ($G_{np}$) is called the magnomechanically induced transparency (MMIT) window. This is shown in Fig. \ref{b}(b). When $g_1 $ is activated, we obtain three windows: the single MIT window depicted in Fig. \ref{b}(a) divides into a double window and MMIT windows, as illustrated in Fig. \ref{b}(c). This result is consistent with that of reference \cite{43}. We can see that the peaks of absorption decreases with the increasing of tunneling coupling strength $f$, thus forming a shallower transparency window. We also note that as the tunneling coupling $f$ increases, the transparency window broadens and becomes deeper. This physical result is very similar to OMIT \cite{Mi}.\\
	In Figs. \ref{b}(d)-\ref{b}(f), we plot the dispersion spectrum Im[$\epsilon_{out}$] of the probe field versus the normalized detuning $\delta/\omega_{p}$ for several values of the cavity-cavity tunneling coupling $f$ and coupling strengths $g_1$, $g_2$ and $G_{np}$ with $G_{au}=0$. A single MIT dispersion spectrum is obtained when only the magnon-photon coupling $(g_2)$ is present, and all other couplings are absent, as shown in Fig. \ref{b}(d). Fig. \ref{b}(e) illustrates the dispersion spectra for the scenarios where $g_1=0$, $G_{np}\neq 0$ and $g_{2}\neq0$. Figure \ref{b}(f) displays the dispersion spectrum of the output field in the presence of the three couplings. We note in these figures that the dispersion spectrum decreases with increasing the tunneling coupling $f$.\\
	\begin{figure} [h!] 
		\begin{center}
			\includegraphics[scale=0.25]{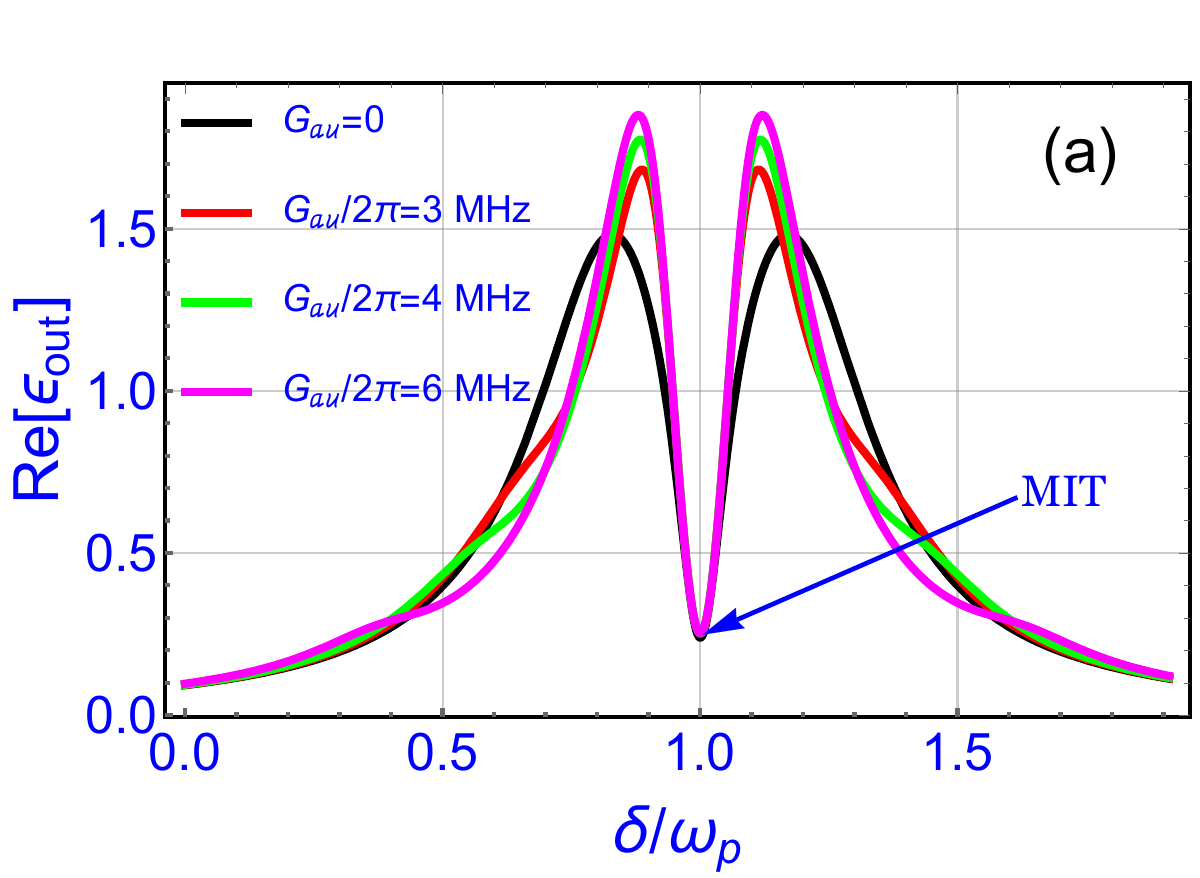}
			\includegraphics[scale=0.25]{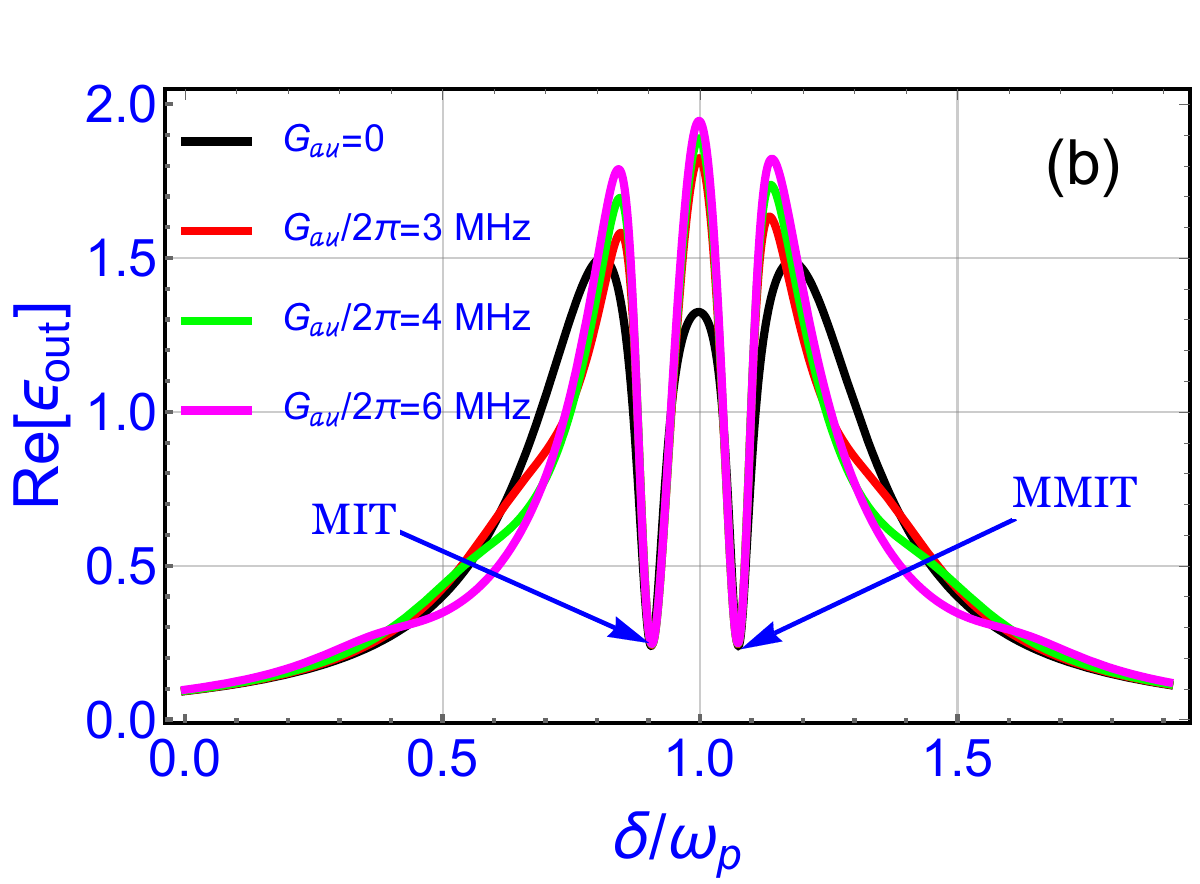}
			\includegraphics[scale=0.25]{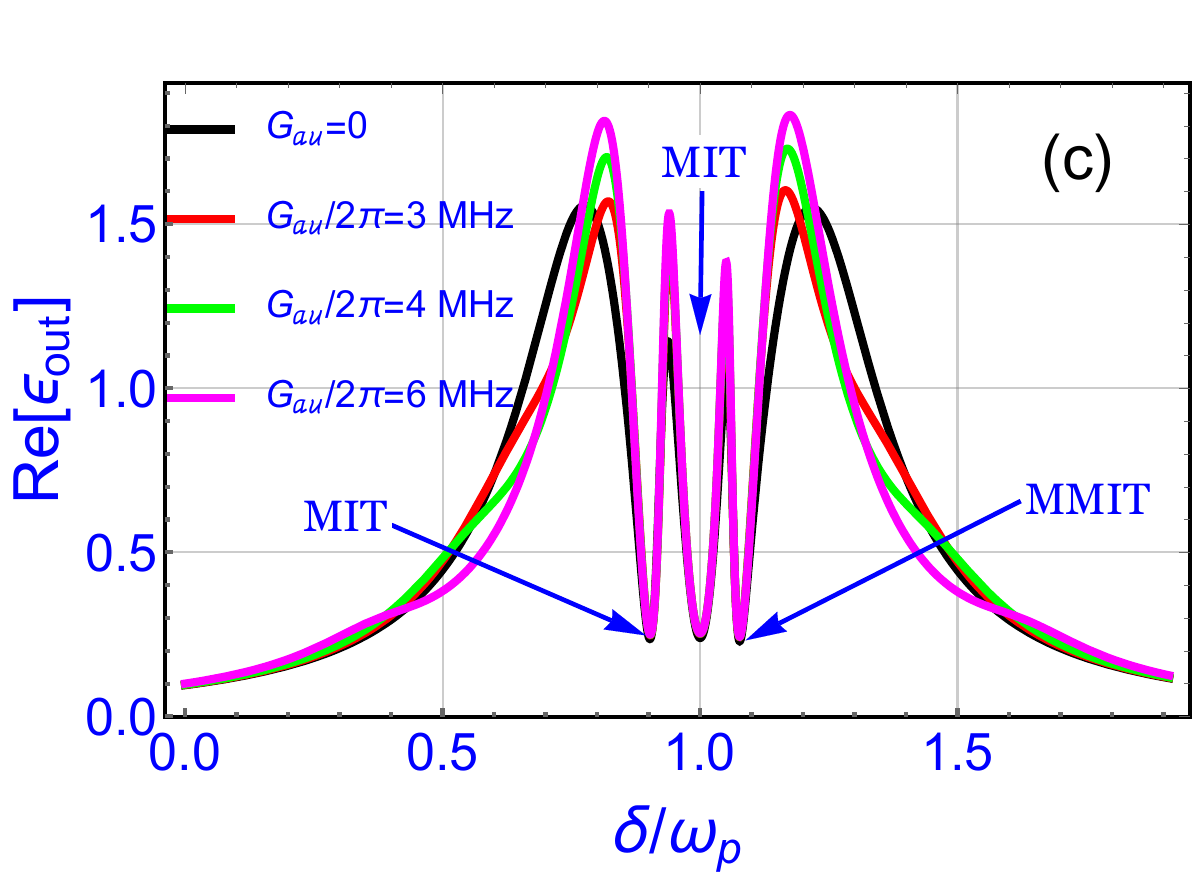}
			\includegraphics[scale=0.25]{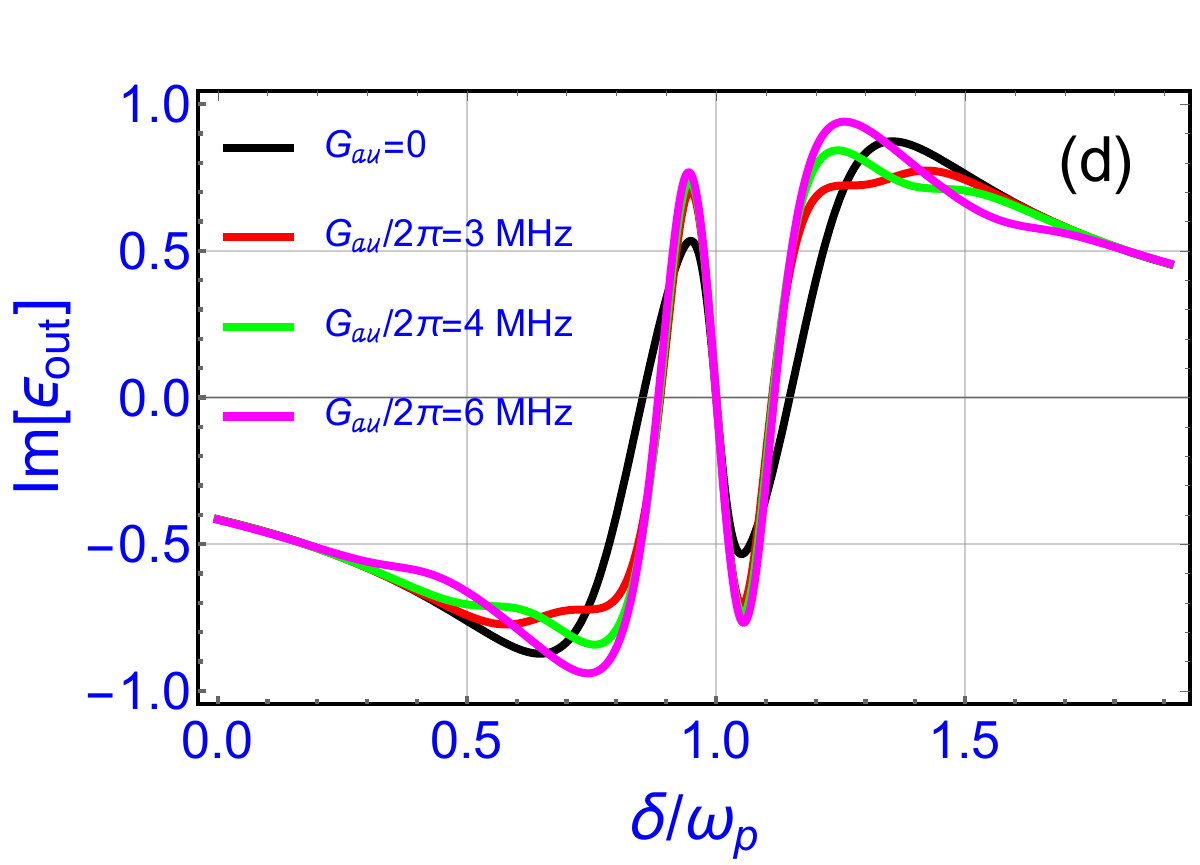}
			\includegraphics[scale=0.25]{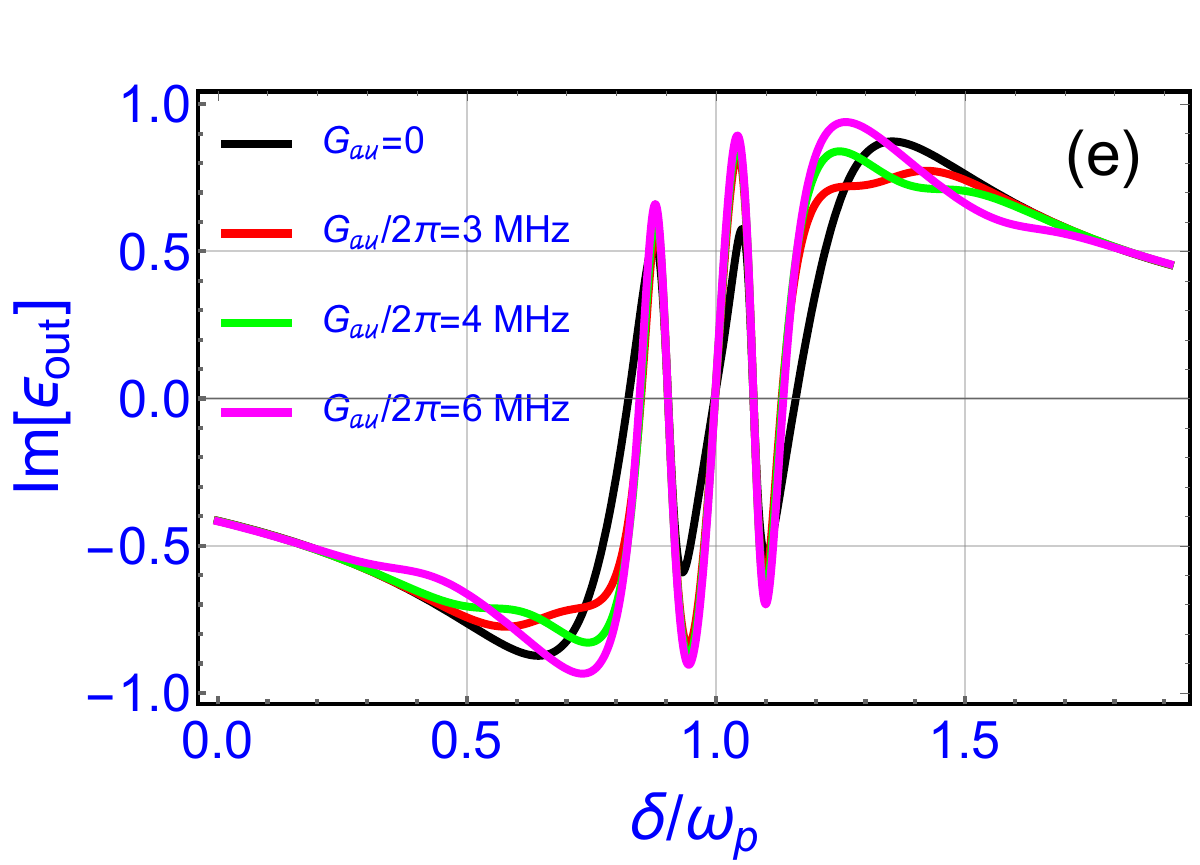}
			\includegraphics[scale=0.25]{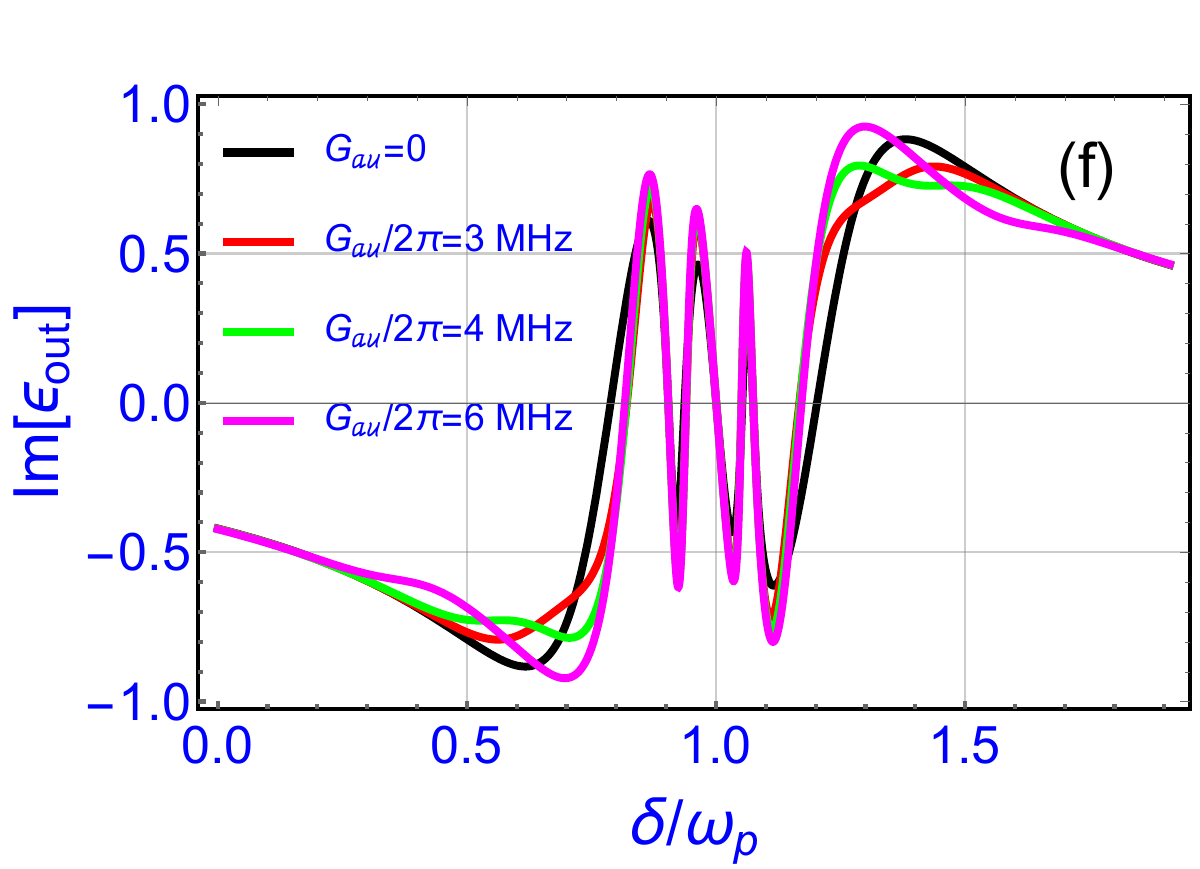}
			\caption{Plot of the real Re[$\epsilon_{out}$] and imaginary Im[$\epsilon_{out}$] parts of the output field as a function of the the probe detuning $\delta/\omega_p$ for different values of the atom-photon coupling $G_{au}$ with $f=0.15\omega_p$. (a)-(d) $g_1=G_{np}=0$ and $g_{2}/2\pi=1.2$ MHz, (b)-(e) $g_1=0$ and $G_{np}/2\pi=g_{2}/2\pi=1.2$ MHz, (c)-(f) $g_1/2\pi=g_{2}/2\pi=G_{np}/2\pi=1.2$ MHz. For additional parameters, see the text.} \label{bb}
		\end{center}
	\end{figure} 
	We plot in Figs. \ref{bb}(a)-\ref{bb}(c) the absorption spectrum Re[$\epsilon_{out}$] of the probe field versus the normalized frequency  $\delta/\omega_p$ for several values of atom-photon coupling  $G_{au}$ and coupling strengths $g_1$, $g_2$ and $G_{np}$ with $f=0.15\omega_p$. We observe one, two, and three windows for the following conditions: when $g_1 = G_{np} = 0$ and $g_2 \neq 0$; when $g_1 = 0$, $G_{np} \neq 0$, and $g_2 \neq 0$; and when $g_1 \neq 0$, $G_{np} \neq 0$, and $g_2 \neq 0$, respectively. Black, red, green, and magenta are the colors depicted in these figures for $G_{au} = 0$, $G_{au} = 3$ MHz, $G_{au} = 4$ MHz, and $G_{au} = 6$ MHz, respectively. We remark that the peaks of absorption increases with the increasing of atom-photon coupling strength $G_{au}$.\\
	In Figs \ref{bb}(d)-\ref{bb}(f), we plot the dispersion spectrum Im[$\epsilon_{out}$] of the probe field as a function of the normalized detuning $\delta/\omega_{p}$ for various values of the atom-photon coupling $G_{au}$ and coupling strengths $g_1$, $g_2$ and $G_{np}$, with $f=0.15\omega_p$. Fig. \ref{bb}(d) shows the single MIT dispersion spectrum in the absence of $g_1$ and $G_{np}$. The dispersion spectra for the cases where $g_1=0$, $G_{np}\neq 0$, and $g_{2}\neq 0$ are shown in Fig. \ref{bb}(e). In Fig. \ref{bb}(f), we depict the dispersion spectrum of the output field when all three couplings are present. We observe in these figures that the dispersion spectrum increases as the atom-photon coupling strength $G_{au}$ increases.
		\begin{figure} [h!] 
		\begin{center}
			\includegraphics[scale=0.25]{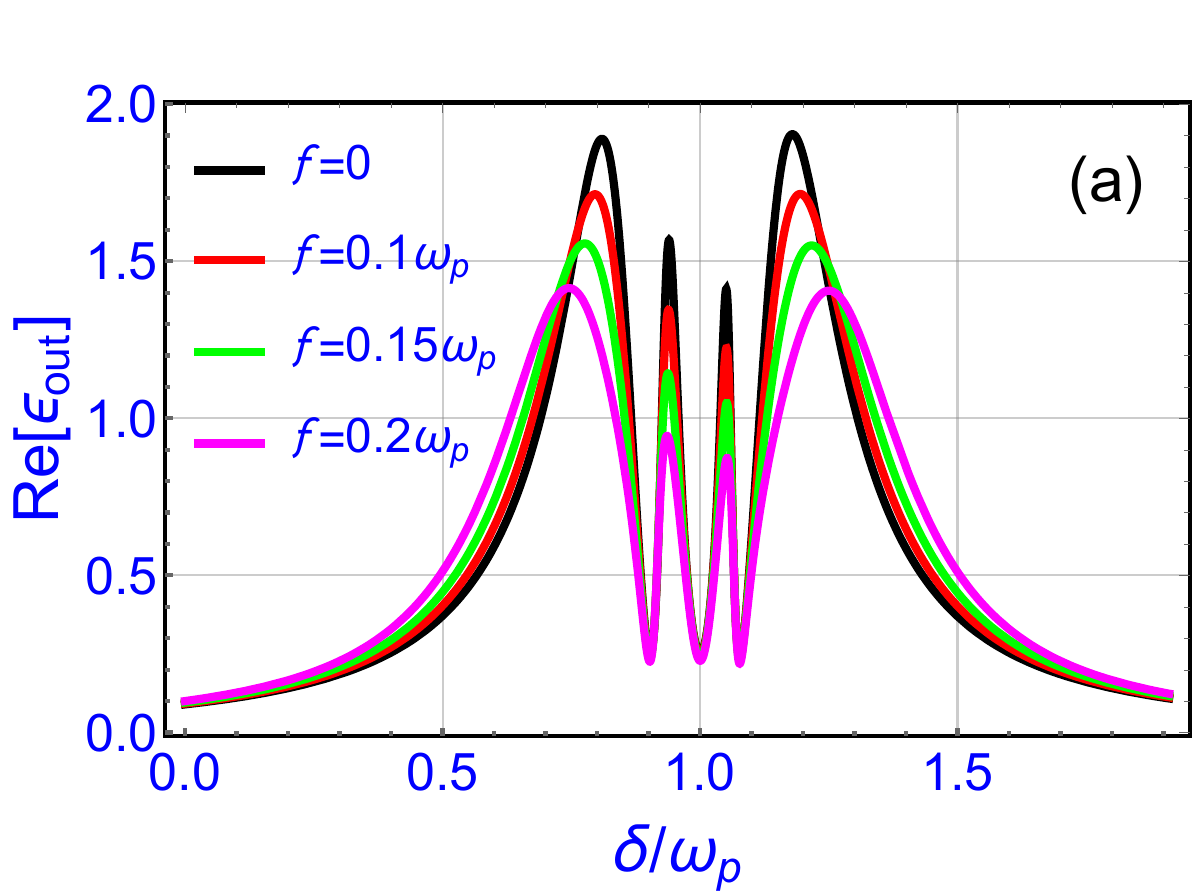}
			\includegraphics[scale=0.25]{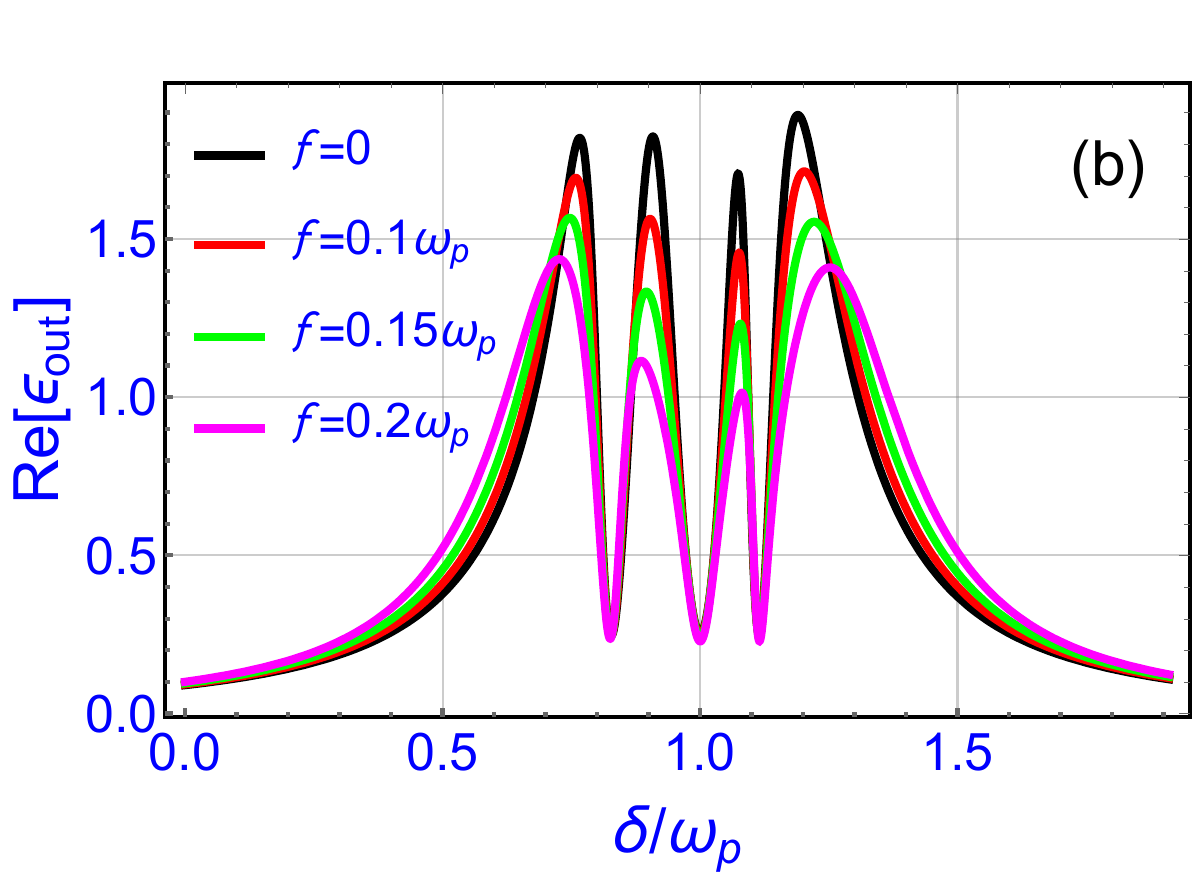}
			\includegraphics[scale=0.25]{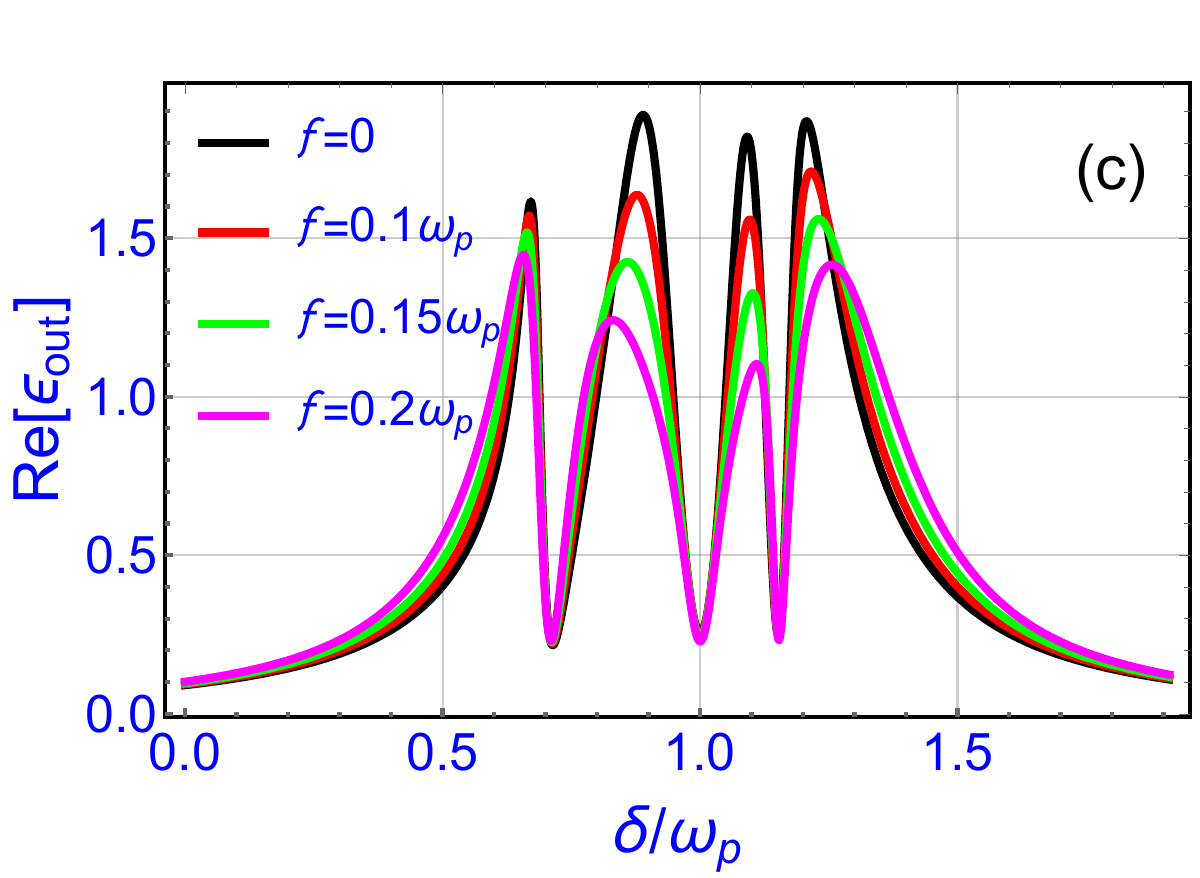}
			\caption{Plot of the real Re[$\epsilon_{out}$] part of the output field as a function of the the probe detuning $\delta/\omega_p$ for different values of the magnetic field $B$ with $f=0$, $f=0.1\omega_p$, $f=0.15\omega_p$ and $f=0.2\omega_p$. (a) $B=0.02$ mT, (b) $B=0.033$ mT, and (c) $B=0.05$ mT. In all panels, $g_{np}/2\pi=1$ mHz, and for additional parameters, see the text. } \label{B}
		\end{center}
	\end{figure} 
In Fig. \eqref{B}, we plot the absorption spectrum of the probe field versus the normalized detuning $\delta/\omega_{p}$ for several values of magnetic field. We remark in these figures that by increasing the magnetic field $B$, the width and peak separations of these windows become larger and more diffuse.
	\section{FANO RESONANCES IN THE OUTPUT FIELD}\label{FFF}
\begin{figure} [h!] 
		\begin{center}
			\includegraphics[scale=0.38]{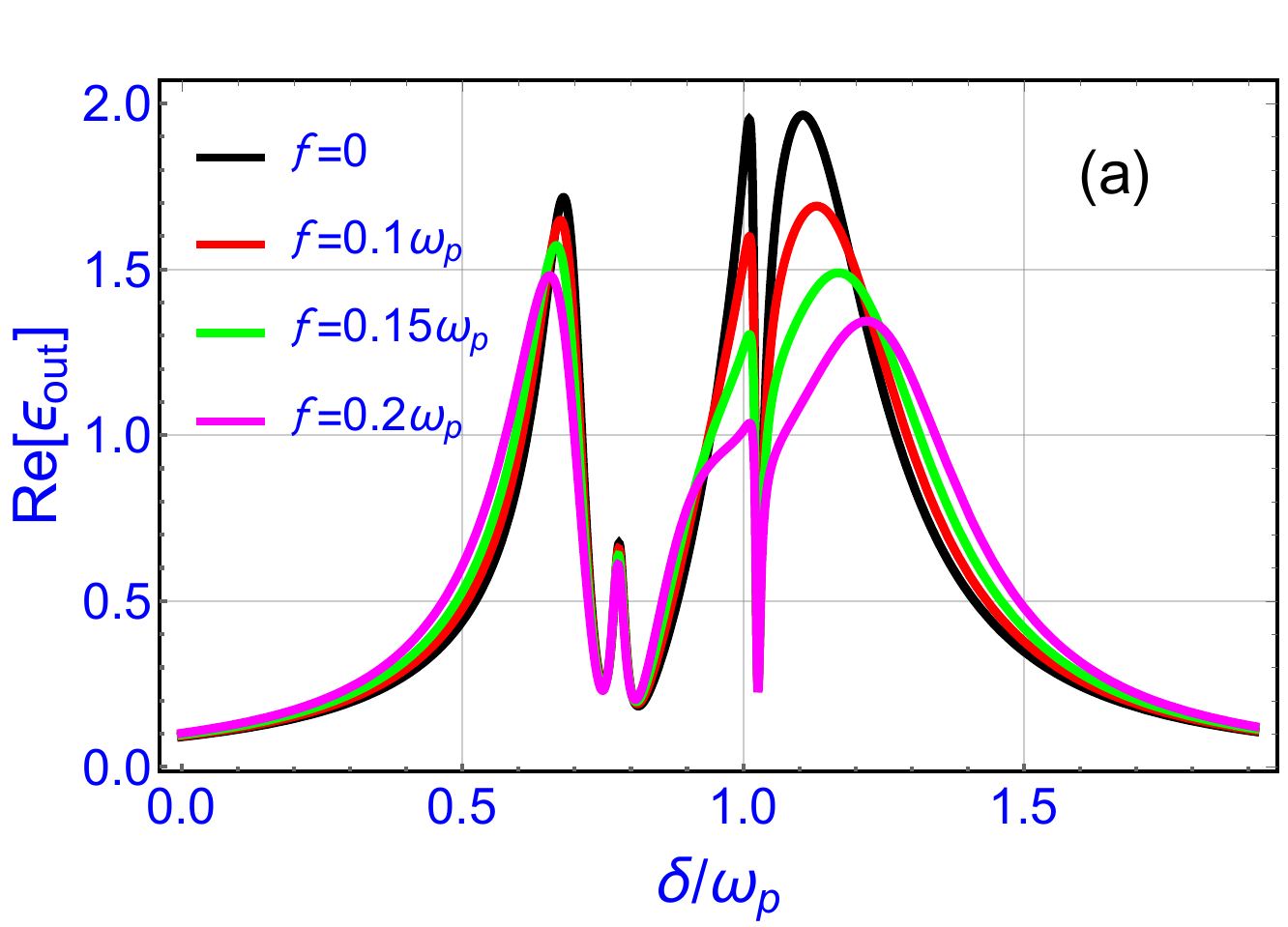}
			\includegraphics[scale=0.38]{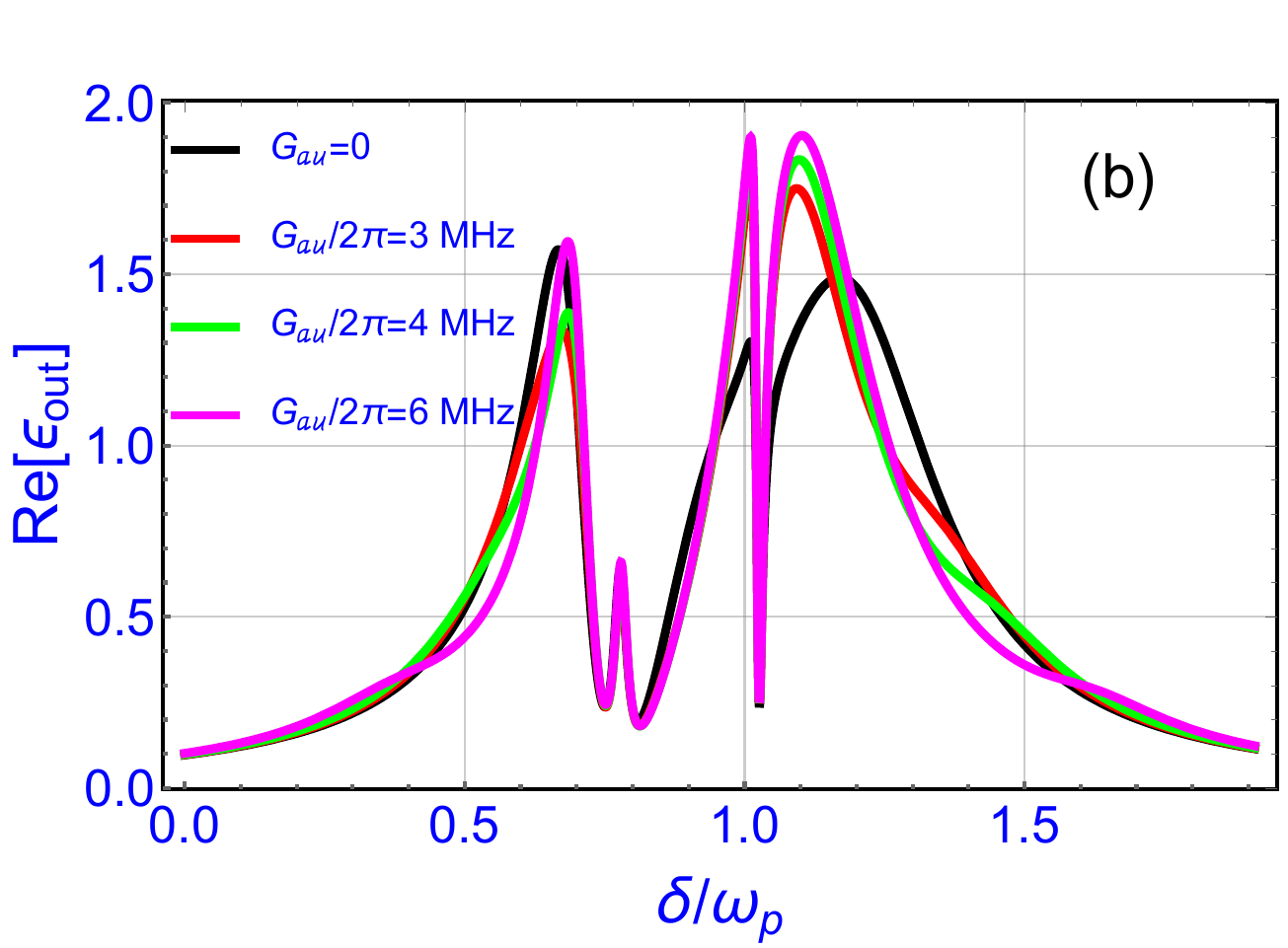}
			\caption{Plot of the Fano lineshapes in the asymmetric real part of the output
				field Re$(\epsilon_{out})$ versus of the normalized detuning $\delta/\omega_{p}$, for several values of (a) tunneling coupling strength $f$, (b) atom-photon coupling $G_{au}$ with $\Delta_{n_{1,2}}=0.8\omega_{p}$. The rest of parameters are defined
				in Sec. \ref{01}.} \label{F}
		\end{center}
	\end{figure}

	In this section, we will explore the Fano resonances present in the output spectrum. The shape of the output spectrum characterized by the Fano resonance is clearly different from the symmetrical resonance curves observed in the EIT, OMIT, MIT and MMIT windows \cite{49,50}. In systems exhibiting optomechanical-like interactions, Fano resonance arises from the existence of non-resonant interactions. In our system, this non-resonant interaction is observed as the magnon mode couples to the phonon mode via an optomechanical-like interaction, where we apply the condition $\Delta_{n_1,n_2} \neq \omega_{p}$. We see in Figs. \ref{F}(a) and \ref{F}(b) asymmetrical Fano shapes with $\Delta_{n_1,n_2} \neq \omega_{p}$ for different $f$ and $G_{au}$, where the absorption spectrum of the output field is shown as a function of the normalized detuning $\delta/\omega_{p}$.
	In Fig. \ref{F}(a), we have fixed the atom-photon coupling at $G_{au} = 0$. Owing to the presence of a non-resonant process $\Delta_{n_1,n_2} =0.8 \omega_{p}$, the absorption spectrum evolves from a symmetric MMIT profile (Fig. \ref{b}(c)) into an asymmetric-window profile for different value of $f$, as illustrated in Fig. \ref{F}(a). From Fig. \ref{F}(b), in the presence of the all couplings and with the presence of the non-resonant process $\Delta_{n_1,n_2} = 0.8 \omega_{p}$, we observe three Fano resonances. This means that the absorption spectrum changes from a symmetrical MMIT profile (Fig. \ref{bb}(c)) to an asymmetrical window profile for different $G_{au}$.
	\section{TRANSMISSION AND SLOW/FAST LIGHT}\label{004}
	In this section, we investigate the impact of the cavity-cavity tunneling coupling $f$ and atom-photon coupling $G_{au}$ on the transmission and group delay of the output field. The transmission of the probe field is defined as the ratio of the amplitudes of the output field to the input field at the probe frequency, as provided by the following expression \cite{41,t1}.\\
	\begin{equation}
		t=\frac{\epsilon_d-2\kappa_aa_{1-}}{\epsilon_d}.
	\end{equation}
	\begin{figure} [h!] 
		\begin{center}
			\includegraphics[scale=0.35]{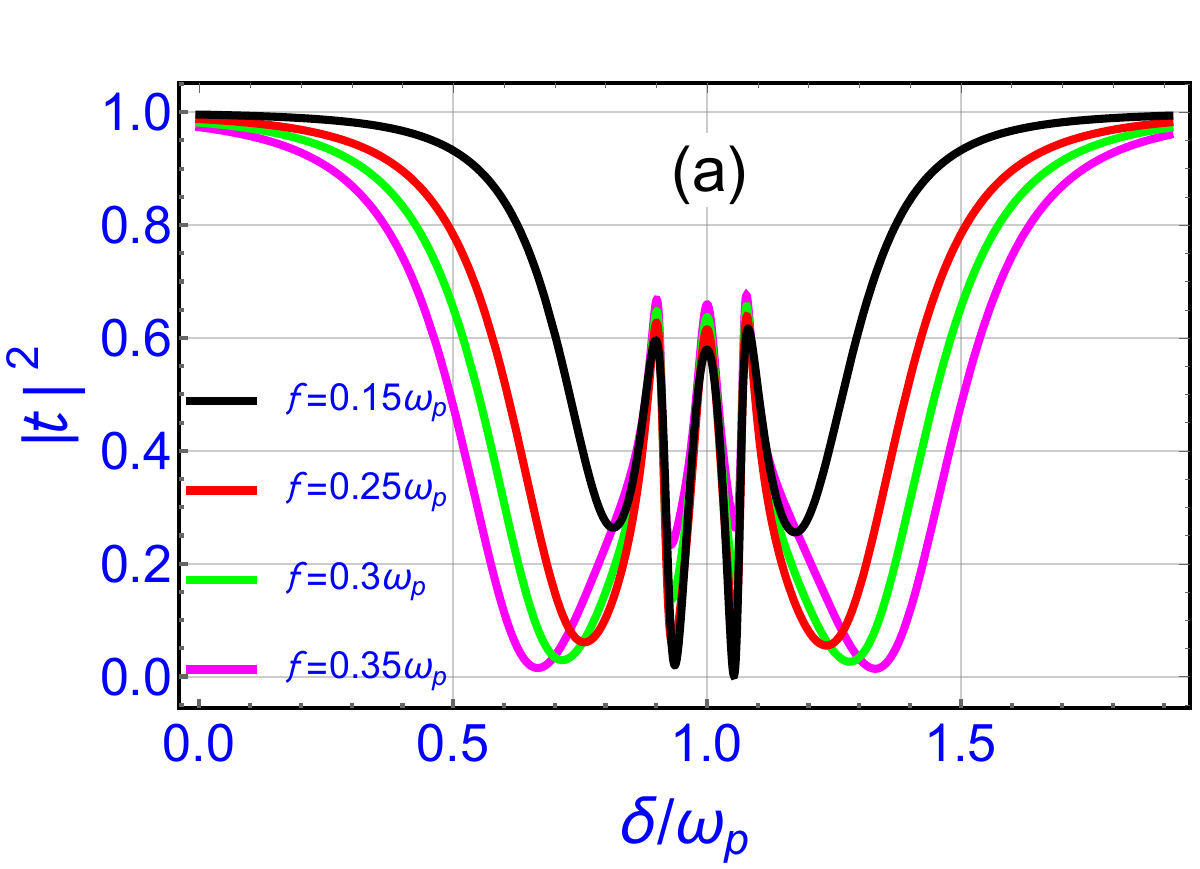}
			\includegraphics[scale=0.35]{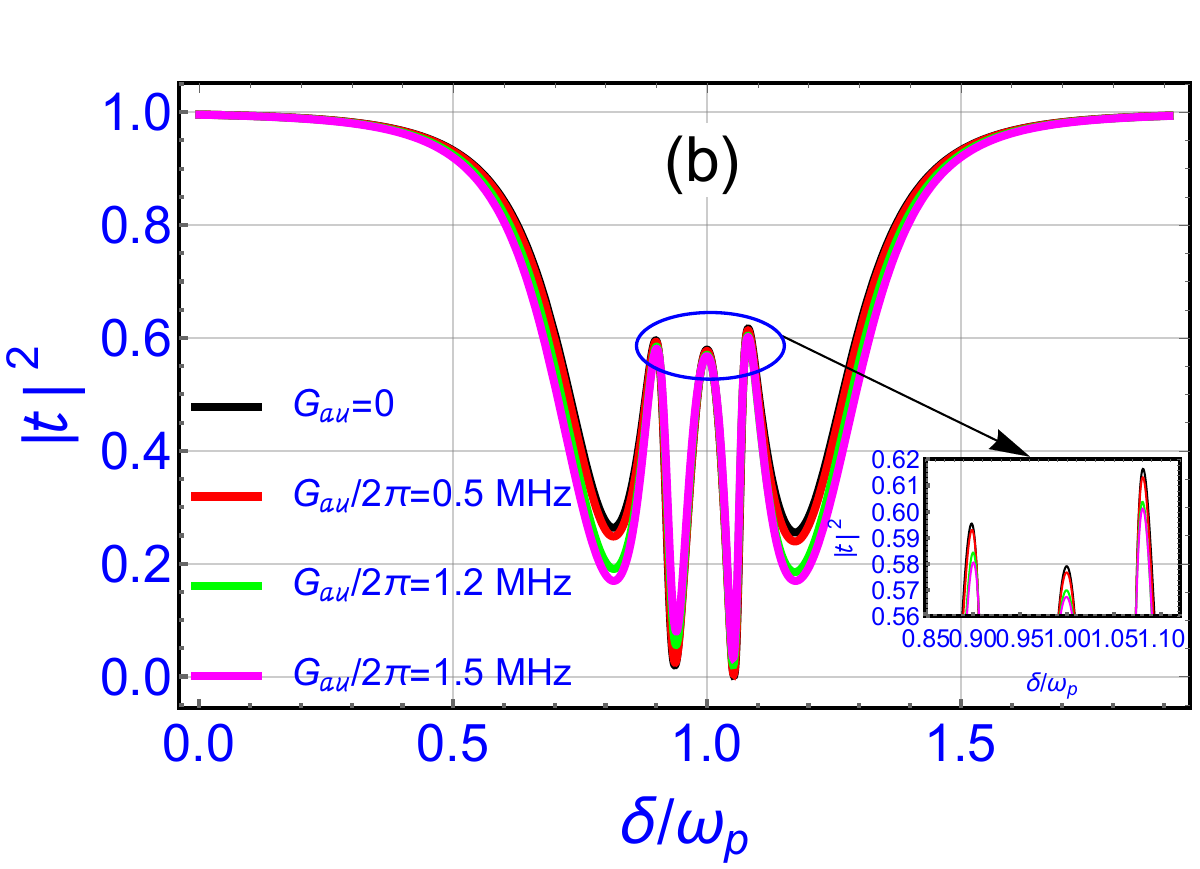}
			\caption{(a) Plot of the transmission $|t|^2$ versus the normalized probe frequency $\delta/\omega_{p}$ for several values of tunneling coupling $f$ with $G_{au}=0$ and $g_1/2\pi=g_{2}/2\pi=G_{np}/2\pi=1.2$ MHz. (b) Plot of the transmission $|t|^2$ versus the normalized probe frequency $\delta/\omega_{p}$ for several values of atom-photon coupling $G_{au}$ with $f=0.15\omega_p$ and $g_1/2\pi=g_{2}/2\pi=G_{np}/2\pi=1.2$ MHz. The rest of parameters are defined in Sec. \ref{01}.} \label{f1}
		\end{center}
	\end{figure} 
	We plot in Fig. \ref{f1}(a) the transmission spectrum $|t|^2$ as a function of normalized detuning $\delta/\omega_{p}$ for several value of the tunneling coupling $f$ with $G_{au}=0$. This figure shows that we have three transparency windows when all couplings are simultaneously nonzero. We remark that the transmission peaks are enhanced when the tunneling coupling $ f $ is increased. For instance, when $ f = 0.2 \omega_p $, the transmission peaks at $\delta/\omega_p \simeq0.90$, $\delta/\omega_p \simeq 1$, and $\delta/\omega_p \simeq 1.07$ are $ |t|^2 = 0.61 $, $ |t|^2 = 0.60 $, and $ |t|^2 = 0.63 $, respectively. In contrast, when $ f = 0.35 \omega_p $, the transmission peaks are $ |t|^2 = 0.67 $, $ |t|^2 = 0.66 $, and $ |t|^2 = 0.35 $.\\
	We plot in Fig. \ref{f1}(b) the transmission spectrum $|t|^2$ versus the normalized detuning $\delta/\omega_{p}$ for different values of the atom-photon $G_{au}$ with $f=0.15\omega_p$ and $g_1/2\pi=g_{2}/2\pi=G_{np}/2\pi=1.2$ MHz. We observe that the transmission peaks diminish as the atom-photon coupling $ G_{au}$ increases. For example, when $ G_{au} = 0 $, the transmission peaks at $\delta/\omega_p \simeq0.90$, $\delta/\omega_p \simeq 1$, and $\delta/\omega_p \simeq 1.07$ are $ |t|^2 = 0.595 $, $ |t|^2 = 0.579 $, and $ |t|^2 = 0.616 $, respectively. In contrast, when $ G_{au}/2\pi = 1.5 $ MHz, the transmission peaks are $ |t|^2 = 0.580 $, $ |t|^2 = 0.567 $, and $ |t|^2 = 0.601 $.\\
	We define a rapid phase dispersion $\Phi=Arg[t]$ within the narrow transparency window. This dispersion can induce a group delay, which is given by the following expression
	\begin{equation}
		\tau  = \frac{{\partial \Phi }}{{\partial {\omega _d}}} = {\mathop{\rm Im}\nolimits} \left[ \frac{1}{t}\frac{{\partial t}}{{\partial {\omega _d}}}\right]. 
	\end{equation}
	A positive group delay $(\tau > 0)$ reflects slow light propagation, whereas a negative group delay $(\tau < 0)$ reflects fast light propagation.\\
	\begin{figure} [h!] 
		\begin{center}
			\includegraphics[scale=0.4]{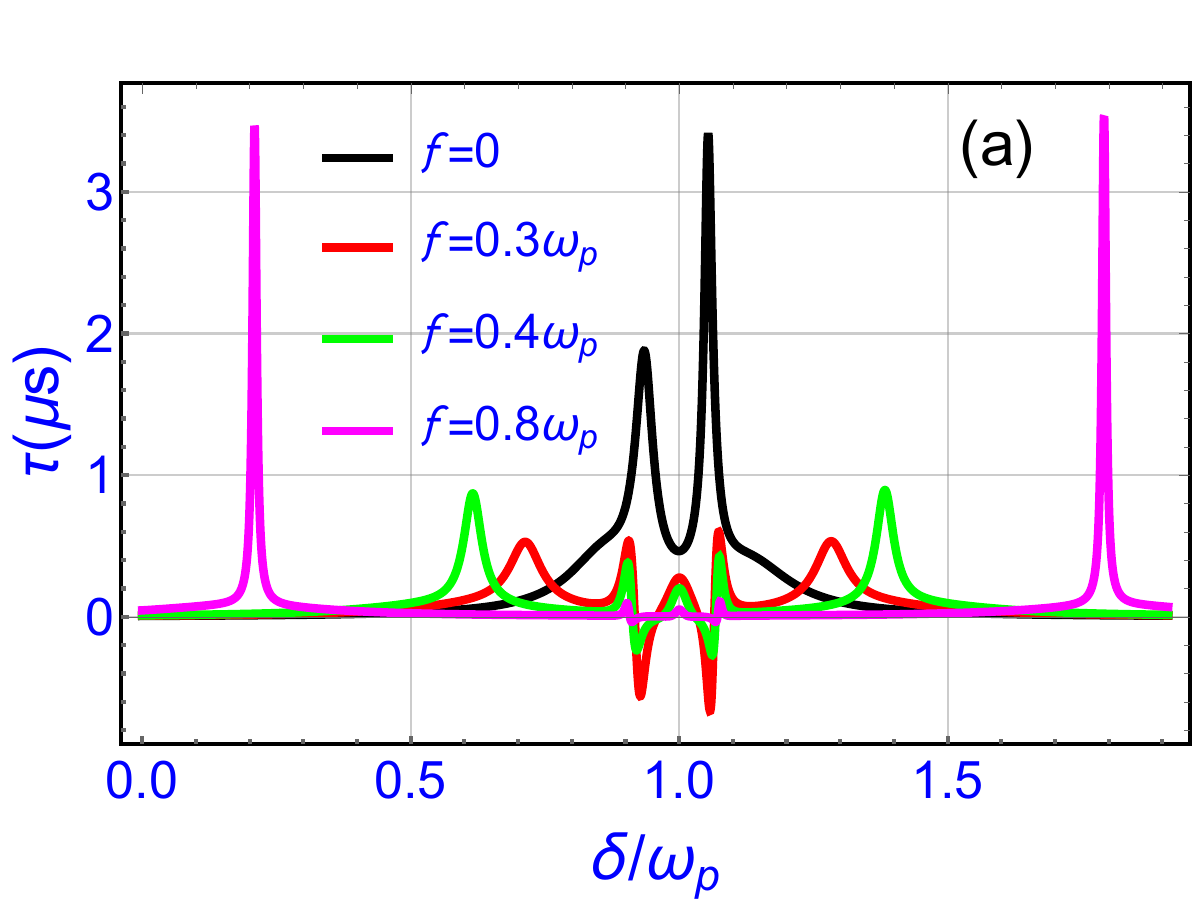}
			\includegraphics[scale=0.4]{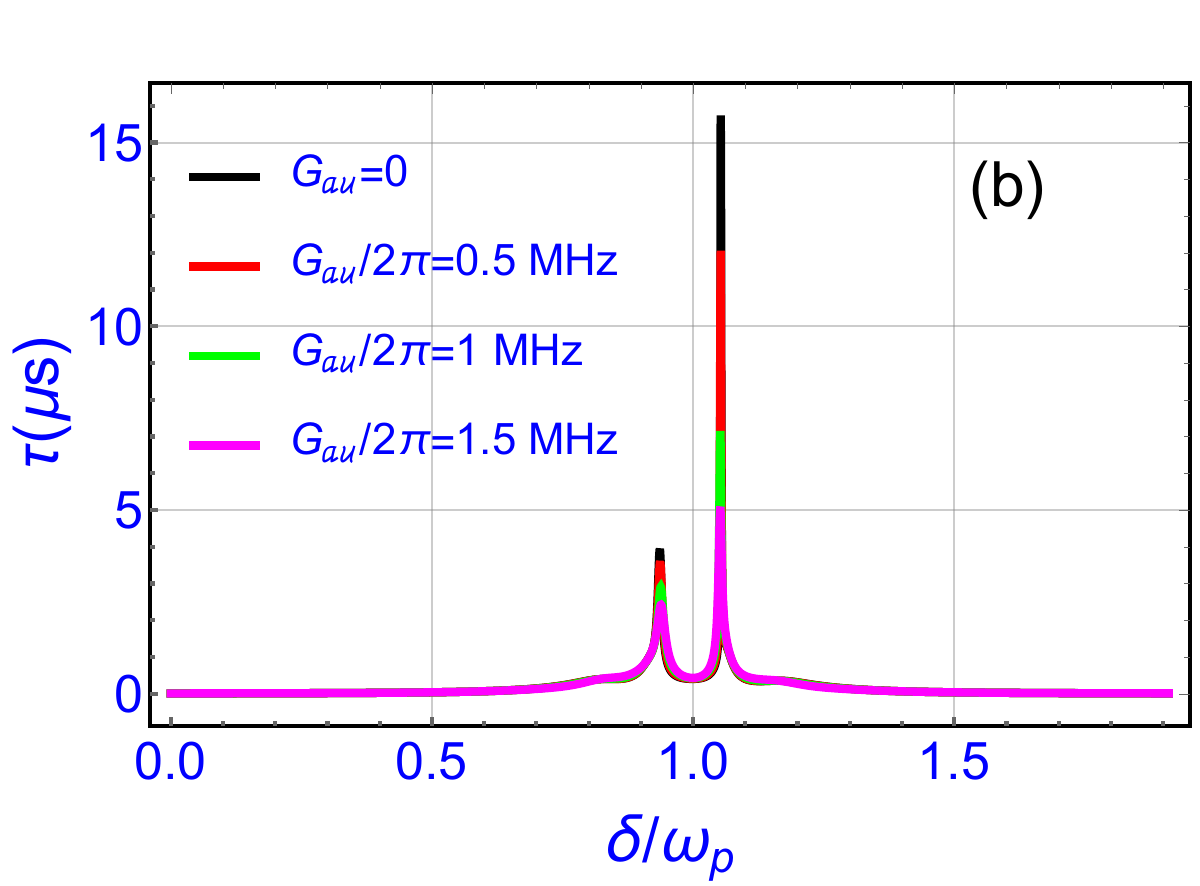}
			\caption{Plot of group delay $\tau$ versus the normalized detuning $\delta/\omega_{p}$ for several values of (a) the tunneling coupling strength $f$ with $G_{au}=0$ and $G_{sp}/2\pi=g_1/2\pi=g_2/2\pi=1.2\,\text{MHz}$, and (b) the atom-photon coupling $G_{au}$ with $f=0.15\omega_p$, $G_{np}/2\pi=g_1/2\pi=g_2/2\pi=1.2\,\text{MHz}$. The rest of parameters are defined in Sec. \ref{01}.} \label{f2}
		\end{center}
	\end{figure} 
	Fig. \ref{f2}(a) shows the group delay $\tau$ as a function of the normalized detuning $\delta/\omega_p$ for different values of the tunneling coupling $f$ with $G_{au}=0$ and $G_{np}/2\pi=g_1/2\pi=g_2/2\pi=1.2\,\text{MHz}$. We observe that when cavity A is not coupled to cavity B, i.e., $f = 0$, the group delay of the output probe field is always positive,  with two peaks observed at $\delta=0.93\omega_p$ and $\delta=1.05\omega_p$. This signifies the slow light effect experienced by the probe field. When the cavity A is coupled with cavity B, i.e., $f = 0.3\omega_p$ , a negative group delay is obtained at $\delta=0.92\omega_p$ and $\delta=1.05\omega_p$, which corresponds to the fast light effect of the output probe field. In contrast, the positive group delay at  $\delta=0.71\omega_p$,  $\delta=0.90\omega_p$,  $\delta=\omega_p$ and  $\delta=1.07\omega_p$, which corresponds to the slow light effect of the output probe field. By increasing the tunneling coupling $ f $, we remark an enhancement of slow light and a decrease in fast light. Therefore, we can enhance the slow light effect and diminish of fast light phenomena by changing the tunneling coupling $f$. By comparing the transmission evolution in Fig. \ref{f1}(a) with Fig. \ref{f2}(a) for $f=0.3\omega_{p}$, it is evident that the peak $(\tau > 0)$ appears at $\omega_d - \omega_0 = 0.71\omega_p$ and $\omega_d - \omega_0 = 1.28\omega_p$. At these points, the transmission transitions from absorption to transmission, leading to an increase in $t$ within this region. In Fig. \ref{f2}(b), we plot the group delay as a function of the normalized detuning $\delta/\omega_p$ with different values of $G_{au}$ with  $f=0.15\omega_p$. We remark that the group delay of the output field is always positive and decreases with increasing the atom-photon coupling $G_{au}$. As a result, we can only observe the phenomenon of slow light. When comparing the variation of transmission in Fig. \ref{f1}(b) with the group delay in Fig. \ref{f2}(b) for $G_{au}/2\pi=1.5$ MHz, we see that the peak $(\tau > 0)$ occurs at $\omega_d-\omega_0=0.93\omega_p$ and $\omega_d-\omega_0=1.05\omega_p$, where transmission is switched from absorption to transmission. The adjustability of the group delay is attributable to the quantum interference that occurs between the probe field and the anti-Stokes field. This indicates that the group delay is subject to dynamic adjustment, and that modulation of the delayed light can be achieved by regulating the cavity-cavity tunnel coupling parameter $f$ and the atom-photon coupling strength $G_{au}$.\\	
	\begin{figure} [h!] 
		\begin{center}
			\includegraphics[scale=0.4]{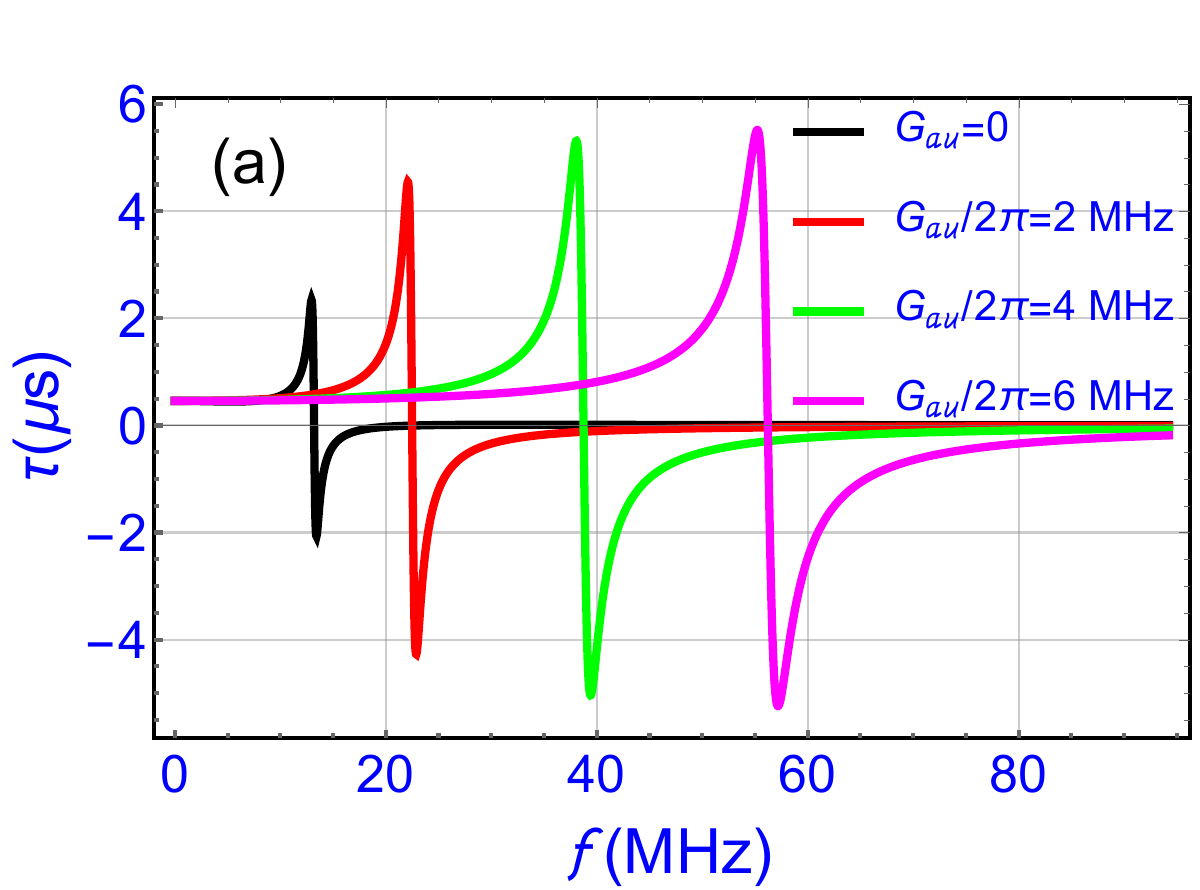}
			\includegraphics[scale=0.4]{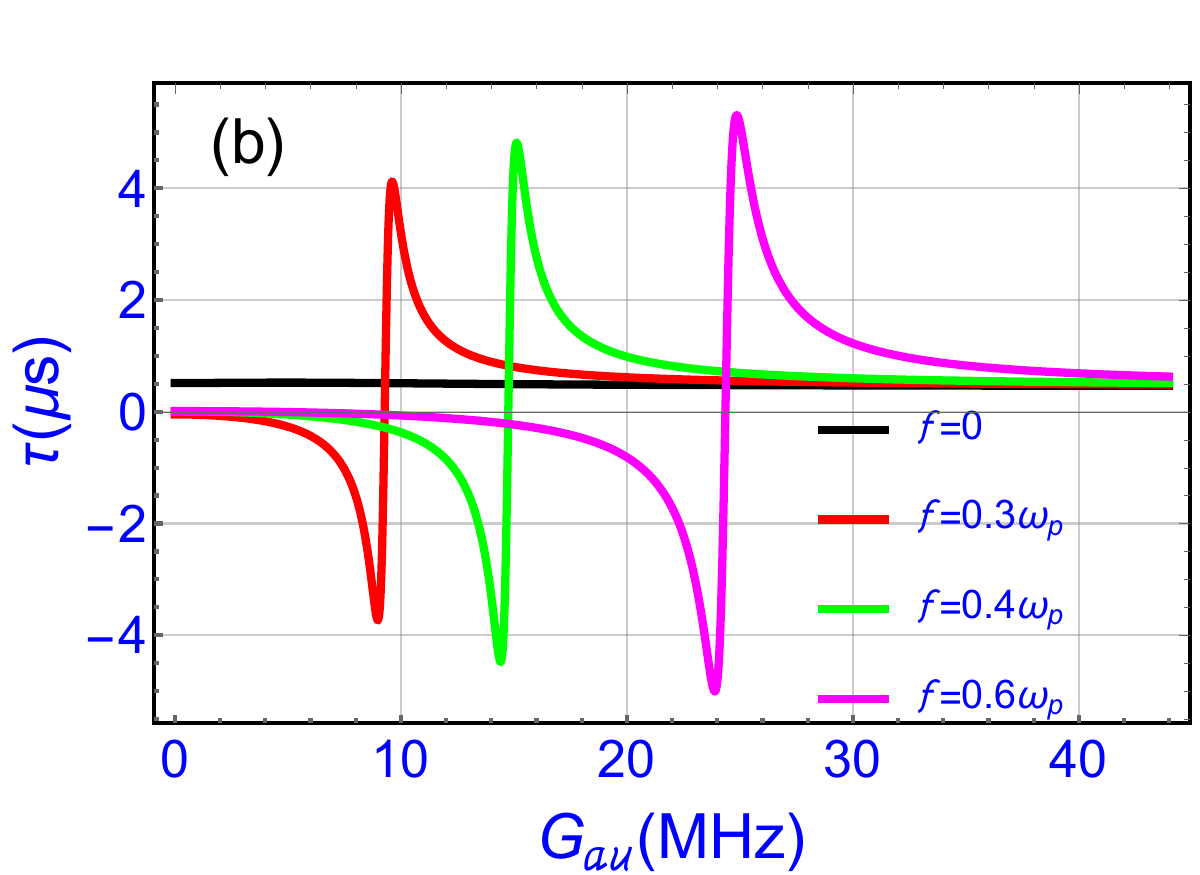}
			\caption{Plot of group delay $\tau$ versus (a) the tunneling coupling strength $f$ for several values of atom-photon coupling strength $G_{au}$ (b) the atom-photon coupling $G_{au}$ for various value of the tunneling coupling $f$ with $g_1=0$, $g_2/2\pi=1.2 \, \text{MHz}$, $G_{np}/2\pi=1.2 \,\text{MHz}$ and $\delta=\omega_p$. The rest of parameters are defined in Sec. \ref{01}.} \label{f3}
		\end{center}
	\end{figure} 
	In Fig. \ref{f3}(a), we plot the group delay $ \tau $ as a function of the tunneling coupling strength $ f $ for various values of the atom-photon coupling strength $ G_{au} $ with $ g_1 = 0 $, $ g_2/2\pi = 1.2 \, \text{MHz} $, $ G_{np}/2\pi = 1.2 \, \text{MHz} $, and $\delta = \omega_p $. We observe the emergence of slow and fast light for a given value of $ G_{au} $. For example, for $ G_{au} = 0 $, the group delay is positive between $ f = 0 $ and $ f = 13.20 \, \text{MHz} $, and negative between $ f = 13.20 \, \text{MHz} $ and $ f = 21.09 \, \text{MHz} $. Additionally, the group delay is improved by adjusting the value of $ G_{au} $, the maximum group delay being reached for higher values of the atom-photon coupling strength.\\
	Fig. \ref{f3}(b) shows the group delay $\tau$ as a function of atom-photon coupling strength $G_{au}$ for different values of the tunneling coupling strength $f$, with $g_1=0$, $g_2/2\pi=1.2 \, \text{MHz}$, $G_{np}/2\pi=1.2 \,\text{MHz}$ and $\delta=\omega_p$. When cavity A is not coupled to cavity B, i.e., $f = 0$, we find that the value of the group delay is positive and fixed at  $\tau=0.5$ $\mu\text{s}$. Conversely, when the tunneling coupling is present, i.e., $f = 0.3\omega_p$, we find two parts of the group delay: a negative part between $G_{au}=0$ and $G_{au}=9.27\, \text{MHz}$, which signifies fast light, and a positive part from $G_{au}=9.27\, \text{MHz}$, which signifies slow light. We remark that the group delay $\tau$ is significantly increased as the tunneling coupling strength $f$ increases.
	\section{FEASIBILITY}\label{00}
	In this study, we investigate the experimental feasibility of our system based on current experimental advancements. The system comprises two interconnected cavities: a primary cavity housing two YIG spheres and an auxiliary cavity containing an atomic ensemble. Recent experiments have demonstrated the interaction between a highly polished single-crystal YIG sphere and a three-dimensional microwave cavity \cite{49}. We also analyzed the parametric regime, which is valid only when the magnon population is significantly smaller than the total spin of the system. Specifically, the average number of magnons, $\langle n^\dag n \rangle$, must be well below $5N$, the total spin of the $Fe^{3+}$ ions in the YIG (where $s = 5/2$) \cite{49}. While the magnomechanical component has been previously demonstrated, the optomechanical aspect has also been experimentally realized \cite{A}. Experimental studies on atom trapping have shown that atoms in their ground and excited states, when properly positioned, remain unaffected by external magnetic fields, as detailed in Ref. \cite{B}. In this work, we adopted parameters from Ref. \cite{43,m13}. Thus, the proposed system consisting of two cavities (the primary cavity with two ferromagnetic yttrium iron garnet (YIG) spheres and the auxiliary cavity with an atomic assembly) connected via photon tunneling appears to be experimentally feasible.\\
	\section{CONCLUSION}
	In conclusion, we have examined the absorption, dispersion, and transmission spectra of a hybrid magnomechanical system. MMIT and MIT can be identified in the absorption and transmission spectra as a result of photon-magnon and phonon-magnon interactions. The cavity-cavity tunneling coupling has the potential to enhance MIT and MMIT. Subsequently, we investigated the impact of the atom-photon coupling on the absorption and transmission spectrum. The results indicate that the transparency windows decrease when the atom-photon coupling is increased. We also studied the influence of the magnetic field on the absorption spectrum. Additionally, the presence of anti-Stokes processes in the system is the cause of the Fano resonance phenomenon, which is thoroughly examined. We have studied the conditions for slow and fast light propagation in our system, which can be regulated by various system parameters. It was demonstrated that the tunability tunneling coupling and atom-photon coupling, has a significant impact on the slow and fast light in a our magnomechanical system. 
	\section*{Acknowledgments}
    M. Amghar expresses gratitude for the financial support he receives from the National Center for Scientific and Technical Research (CNRST) under the “PhD-ASsociate Scholarship-PASS” program.

\end{document}